# A sequential approach to calibrate ecosystem models with multiple time series data

Ricardo Oliveros-Ramos[a,b,c], Philippe Verley[b,c] and Yunne-Jai Shin[b,c]

[a] Instituto del Mar del Perú (IMARPE). Gamarra y General Valle s/n Chucuito, Callao, Perú.

[b] Institut de Recherche pour le Développement (IRD), UMR EME 212, Avenue Jean Monnet, CS 30171, 34203 Sète Cedex, France.

[c] University of Cape Town, Marine Research (MA-RE) Institute, Department of Biological Sciences, Private Bag X3, Rondebosch 7701, South Africa.

Corresponding author: Ricardo Oliveros-Ramos: ricardo.oliveros@gmail.com

Abstract

Ecosystem approach to fisheries requires a thorough understanding of fishing impacts on ecosystem status and processes as well as predictive tools such as ecosystem models to provide useful information for management. The credibility of such models is essential when used as decision making tools, and model fitting to observed data is one major criterion to assess such credibility. However, more attention has been given to the exploration of model behavior than to a rigorous confrontation to observations, as the calibration of ecosystem models is challenging in many ways. First, ecosystem models can only be simulated numerically and are generally too complex for mathematical analysis and explicit parameter estimation; secondly, the complex dynamics represented in ecosystem models allow species-specific parameters to impact other species parameters through ecological interactions; thirdly, critical data about non-commercial species are often poor; lastly, technical aspects can be impediments to the calibration with regard to the high computational cost potentially involved and the scarce documentation published on fitting complex ecosystem models to data. This work highlights some issues related to the confrontation of complex ecosystem models to data and proposes a methodology for a sequential multi-phases calibration of ecosystem models. We first propose two criteria to classify the parameters of a model: the model dependency and the time variability of the parameters. Then, these criteria and the availability of approximate initial estimates are used as decision rules to determine which parameters need to be estimated, and their precedence order in the sequential calibration process. The end-to-end (E2E) ecosystem model ROMS-PISCES-OSMOSE applied to the Northern Humboldt Current Ecosystem is used as an illustrative case study. The model is calibrated using a novel evolutionary algorithm and a likelihood approach to fit time series data of landings, abundance indices and catch at length distributions from 1992 to 2008.

## 1. Introduction

The implementation of an ecosystem approach to fisheries not only requires a thorough understanding of the impact of fishing on ecosystem functioning and of the ecological processes involved, but also quantitative tools such as ecosystem models to provide useful information and predictions in support of management decision. Yet, the use of ecosystems models as decision making tools would only be possible if they are rigorously confronted to data by means of accurate and robust parameter estimation methods and algorithms (Bartell 2003). In many respects, the calibration of ecosystem models is a complex task. In particular, the dynamics represented in ecosystem models allow species-specific parameters to have an impact on one another through ecological interactions, which results in highly correlated parameters. In addition, critical information and observations on non-commercial species can be missing or poor. Furthermore, the large number of parameters and the long duration of the simulations can be an obstacle to calibrate a model. These diverse reasons hampered the development of flexible and generic calibration algorithms and methodology for ecosystem models, and only sparse documentation has been produced on fitting complex models (Bolker et al. 2013).

Given that calibration of complex ecosystem models requires a lot of data and potentially involves a large number of parameters to be estimated, common practice in the field has been to i) reduce the number of parameters to be estimated by directly using estimates from other models (Marzloff et al. 2009, Lehuta et al. 2010) or available for similar species or ecosystems (Bundy 2005, Ruiz and Wolff 2011), ii) use outputs from other models as data to calibrate the model (Mackinson and Daskalov 2007), or iii) use both strategies (Shannon et al. 2003, Guénette et al. 2008, Friska et al. 2011, Travers-Trolet et al. 2013). These different strategies expedite the calibration of complex models while attempting to synthesize the maximum of available information. However, since the borrowed parameters and outputs rely on different model assumptions, they may lead to inaccuracy and inconsistency in parameter estimation by trying to reproduce other models' dynamics."

The defect of these common practices can be overcome by implementing a multiple-phase calibration approach (Nash and Walker-Smith 1987, Fournier et al. 2012). In this multiple-phase approach, some parameters can be fixed at initial values obtained from independent data, other models or expertise (Nash and Walker-Smith 1987). (Nash and Walker-Smith 1987). In particular, assigning initial guess values for completely unknown parameters before proceeding to a full calibration of all parameters can ease the estimation of model parameters (Nash and Walker-Smith 1987, Fournier et al. 2012). This multiple-phase calibration approach is supported by some optimization softwares, like specialized R packages or the AD Model Builder software (Bolker et al. 2013). However, it is difficult to find in the literature a clear roadmap or strategy to guide the users and help them to determine what parameters should be estimated in the successive phases. It appears that the final organization of the calibration phases is most often an empirical process and is the result of trials and errors in the calibration procedure (Fournier 2013).

The objective of this paper is to highlight some issues related to the confrontation of complex ecosystem models to data and propose a methodology to a sequential calibration of ecosystem models, illustrating it with the calibration of the ecosystem model OSMOSE (Shin and Cury 2004, Travers et al. 2009) applied to the northern Humboldt Current Ecosystem. The first important step in a calibration is to be able to categorize the parameters of a model. To do so, we propose two criteria: the model dependency and the time variability of the parameters. Then, we use these criteria and the availability of initial guess values of the parameters to determine which parameters need to be estimated, and their precedence in the sequential calibration process. We finally compare our sequential approach with the results of a single step calibration of all parameters.

## 2. Material and methods

### *2.1.* Parameterization and calibration

#### *2.1.1 Types of parameters*

Several classifications of model parameters can be found in the literature (e.g. Jorgensen and Bendoricchio 2001) according to different criteria and for different purposes. In this work, we classified the parameters according to two criteria: 1) the dependence of the parameter on the model structural assumptions, and 2) the time variability of the parameter in relation to its use in the model. The categorization of the parameters is defined as follows.

**Model dependency:** Parameters are considered to be model-dependent when their values can vary between models due to different model structures or assumptions. For example, fishing mortality can be categorized as being model-dependent, because it depends on the value of natural mortality, structural equations of the fishing process and assumptions on the selectivity or seasonal distribution of fishing effort. On the contrary, model-independent parameters can be estimated directly from data and observations by simple models or theoretical relationships. For example, parameters for the length-weight relationships or for the von Bertalanffy growth function can be considered independent of the overarching ecosystem model structures and assumptions.

**Time variability:** Some parameters of the model are expected to have temporal variability at the time scale of the model and the data. For example, fish larval mortality rates which determine the fish annual recruitment success and which are related to environmental conditions are expected to vary annually. Other parameters of an ecosystem model are not expected to have significant temporal variability at the time scale of the model and the data time series, for example the parameters of predators' functional response.

The classification of the parameters in terms of model dependency is necessary in order to avoid the misleading use of parameters' values which have been estimated in other models and not directly from observations. If some parameters are fixed at values inconsistent with the model structure currently used to fit the data, the estimates of other parameters obtained from the calibration can be highly uncertain and only artifacts to fit the data. This can also impede the convergence of the objective function and lead to a calibration failure (Gaume et al. 1998, Whitley et al. 2013).

The classification in terms of temporal variability can be more arbitrary since many parameters (especially the ones characterizing the populations) are expected to vary with time. The cutoff we propose for a parameter to be considered as time-varying results from the following considerations: i) the identification of a process leading to such time variability, ii) the existence of theoretical assumptions about the importance of such process in the dynamics of the modeled ecosystem, iii) the non-explicit representation of the process in the model, and iv) the significance of the time variability compared to the time scale of the model and the length of the data time series. Some parameters can be assumed to be constant at shorter time scales (e.g. a few years) but can exhibit variability at longer time scales (e.g. several decades). For example, the length at maturity for a given species can decrease in response to heavy fishing (Shin et al. 2005), but can be considered as constant in the model for periods short enough, or if the variability is not considered to cause significant changes in the functioning of the multispecies assemblage.

Despite the apparent dichotomous classification presented, the degree of model-dependency and temporal variability in the parameters can vary, and a qualitative classification of the parameters should be attempted. In the OSMOSE ecosystem model, such classification could be proposed for the parameters characterizing modelled multispecies fish assemblages (Figure 1; see Appendix A for details about the parameterization of OSMOSE).

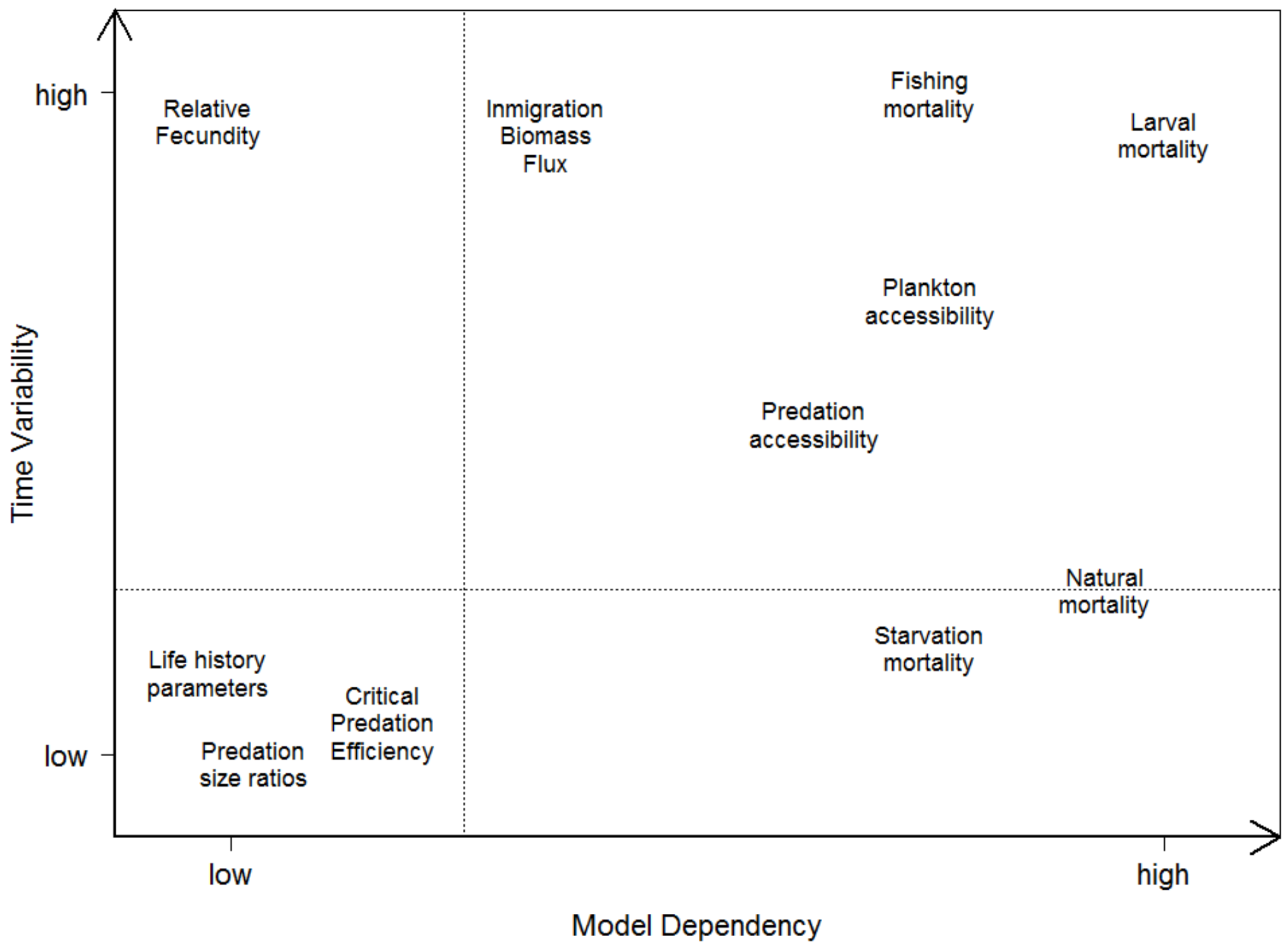

Figure 1. Proposed classification of OSMOSE model parameters depending on the time variability and the model dependency criteria.

## *2.2 Approach for the sequential calibration*

### *2.2.1 Progressive time resolution of the parameters*

The number of parameters to be estimated in a model can be high, particularly when time-varying parameters are included, so that fitting the model to data can be challenging (e.g. see Schnute 1994). Additionally, the way a model is parameterized will define the objective function to be optimized to estimate the parameters; just by rescaling or transforming the parameters this objective function can be changed and the parameter estimation process can be improved (Bolker et al. 2013).

There are several ways to model the time variability in the parameters, taking into account the assumed shape of the variability and the degree of time resolution one wants to introduce (see Megrey 1989, Methot and Wetzel 2013 for examples in fishery models). However, in practical terms, there is a limit in the number of parameters that can be estimated, which depends on the quality of the available data to estimate such parameters. The data must provide information on the time variability of the parameters at the right resolution, otherwise the risk is high to end fitting the noise in data or simply failing in the parameter estimation. This means the decision to keep a parameter constant or to model its time variability has to take into account both the complexity of the parameter estimation and the quality of the data used for the calibration of the model (Jorgensen and Bendoricchio 2001).

We considered different models to represent the temporal variability in the parameters (Appendix B, Table B.1, Figure B.1), where the variability can be split into three components: the mean value of the parameter, the high frequency variability (seasonal or non-periodical) and the low frequency variability (interannual). This type of parameterization allowed us to define several nested models, i.e. models which can gradually be complexified from the simplest models' parameterization (setting to zero the random effects in the yearly component for the time-varying parameters) to full consideration of low and high frequency variability. For example, the mean value of time-varying parameters over the time series should be estimated in priority, with the interannual deviations fixed to zero.

If the different components of time variability in a given parameter can be introduced progressively, the final parameter estimation can be improved by providing good initial values for the more important variability components after preliminary calibration of simpler versions of the model. We therefore propose a general calibration strategy that models the time-varying parameters such that the several components of variability are independent and can be nullified by fixing some parameters to constant values (nested models).

#### *2.2.2. Calibration in multiple-phase*

According to the parsimony principle, given models with similar accuracy, the simplest model is the best compromise with regard to the available data. In particular, for nested models, the complexity of a model is directly related to the number of parameters to estimate, so the parsimony principle implies to estimate the lowest number of parameters as possible. On the other hand, it is possible to increase the goodness of fit of the model by increasing the number of parameters, but this can lead to overfitting (Walter and Pronzato 1997, Bolker 2008). However, there is no way to know a priori if all parameters will be identifiable, i.e. if we can estimate them properly from the available data.

Based on the criteria of time variability and model dependency of the parameters, we propose a set of rules for a hierarchical approach to select the parameters to estimate in a model and the order at which the parameters should be estimated in the different phases of the calibration. Also, we propose some criteria to design nested models for taking into account time variability by using simple time series models which allow to assess the usefulness of the additional temporal parameters introduced in the model calibration.

The first rule relates to the model dependency of the parameters. Apart from the time variability, parameters with low model dependency should be assigned values directly from observations, simple models or from dedicated designed experiments. On the contrary, parameters with high model dependency should always be estimated through model calibration, because even though the theoretical meaning of the parameters is not necessarily model-dependent, different model structures will introduce differences in the actual meaning and value of the parameters within the model.

The second rule relates to the time variability of the parameters. Along with the decision on what components of the variability need to be included (e.g. seasonal or interannual), it is also necessary to assume which time component is more important to explain the total variability of the parameter. These choices allow to progressively increase the variability of the time-varying parameters during different phases of the calibration. The order of the calibration phases is different depending on whether the seasonal component (Figure 2 A and C) or the interannual component (Figure 2B and D) dominates the temporal dynamics.

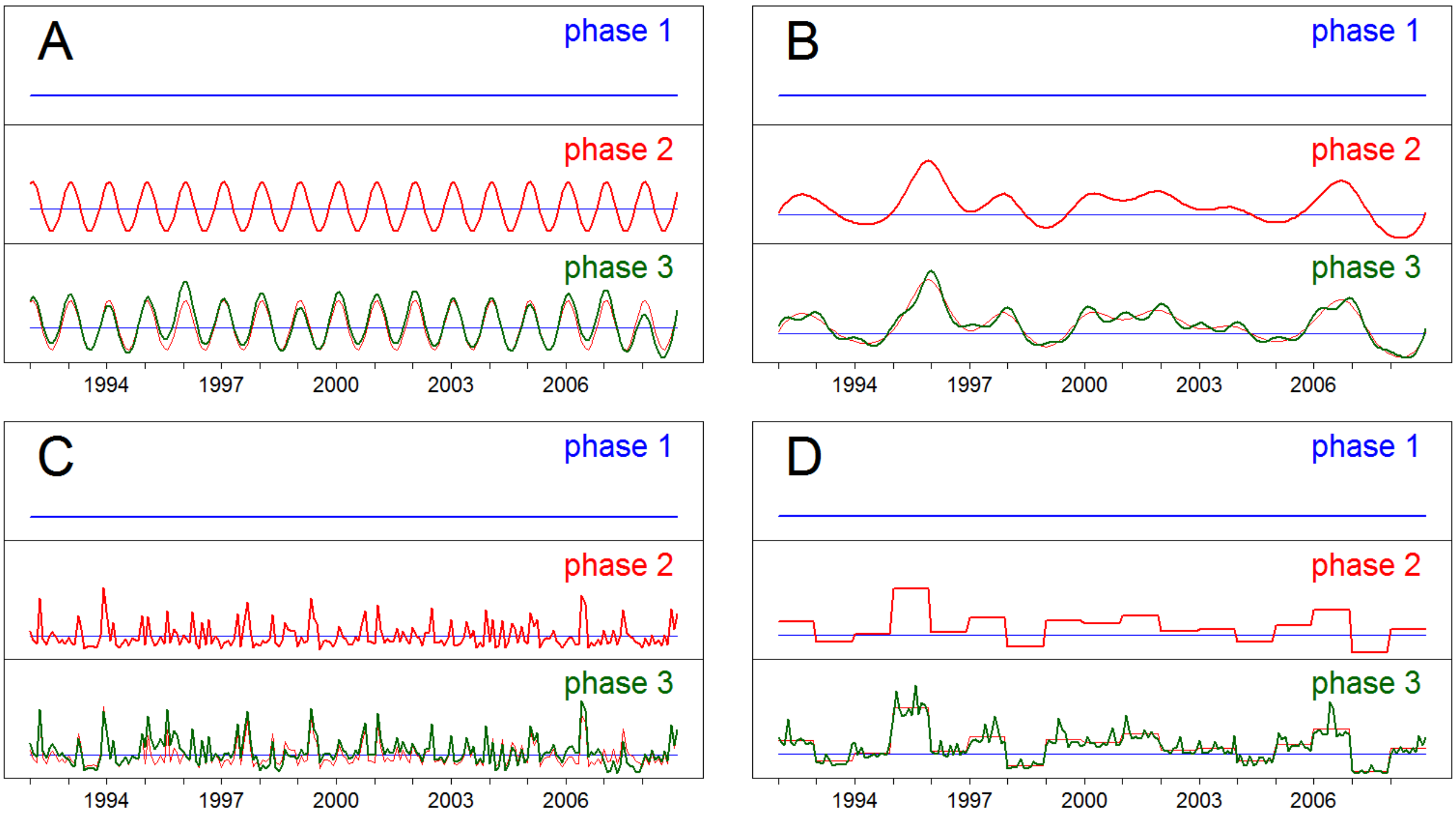

Figure 2. Progressive increase in the time variability of parameters in the case of a 3-phases model calibration. The A and C plots show predominance in the seasonal variability,

while interannual variability is the dominant signal in plots B and D plots. The upper plots correspond to type-A models and the lower plots to type-B models described in the appendix B.

A third rule relates to the availability of initial estimates for the parameters to calibrate. Even in case of model-dependency, parameters taken from other models can be used to start the calibration. For time-varying parameters, the deviations estimated from other models (e.g. single species models) or approximate relationships can be used as proxies of the interannual variability if they are estimated from the same dataset used for the current calibration. Another alternative is to use parameters estimated from models with a similar structure, such as previous versions (likely simpler) of the same model (e.g. a steady-state one). It is important to note that the proxies or initial estimates will only be used to start the calibration in a given parameter space but the parameters will be fully estimated during the calibration, which means the final estimates can be far from the initial values (Sonnenborg et al. 2003).

Considering all these rules together, we propose a hierarchy in the order of parameters' estimation using a sequential calibration approach (Figure 3).

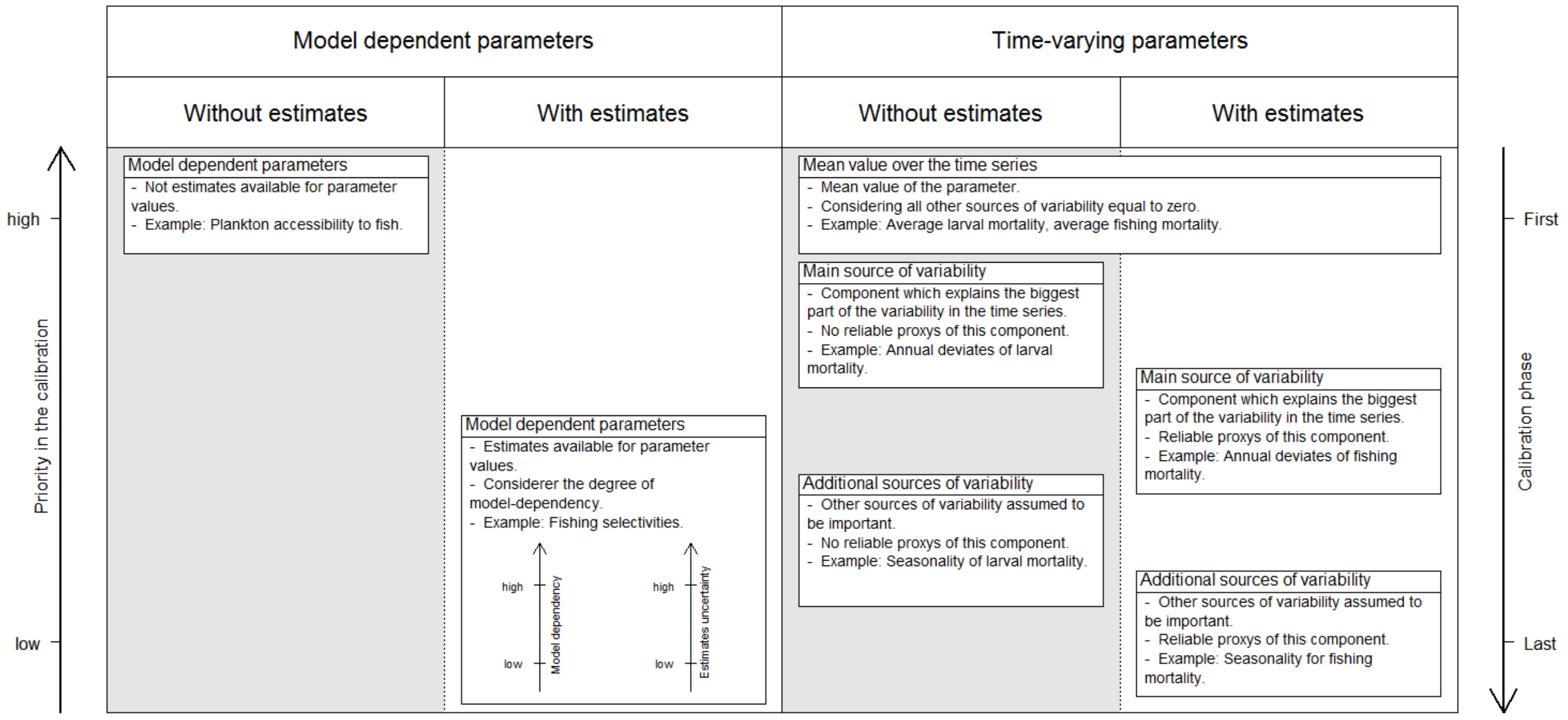

Figure3. Proposed hierarchy in the order of parameters' estimation using a sequential calibration approach.

If we consider the initial phase of the calibration with some parameters fixed as a way to improve the final calibration, it is possible to make changes in the objective function across the different phases of the calibration (Fournier 2013). However, by keeping constant the objective function and running a full optimization at each phase, it is possible to analyze the improvement in the fitting process as a result of the increased complexity of the calibration. It therefore allows to test the usefulness of the additional parameters and to perform model selection by detecting which parameters do not improve the data fitting.

### 2.3. Case model: OSMOSE for Northern Humboldt Current Ecosystem

To illustrate our calibration methodology, we applied the OSMOSE model to the Northern Humboldt Current Ecosystem (NHCE) inhabited by the main stock of "anchoveta" or Peruvian anchovy (*Engraulis ringens*). As the paper focuses on the calibration methodology, We do not present the OSMOSE model in detail but provide key information in Appendix A. Details of the OSMOSE model can be found in Shin and Cury (2001, 2004) and Travers et al. (2009, 2013), and the application to the Humboldt ecosystem is detailed by Oliveros et al. (in prep.) as well as in Appendix A. OSMOSE is a multispecies individual-based model (IBM) which focuses on high trophic level (HTL) species. This model assumes size-based opportunistic predation based on the spatial co-occurrence of a predator and its prey. It represents fish individuals grouped into schools, and models the major processes of fish life cycle (growth, predation, natural and starvation mortalities, reproduction and migration) and the impact of fisheries. The modeled area ranges from 20°S to 6°N and 93°W to 70°W covering the extension of the Northern Humboldt Current Ecosystem and the Peruvian Upwelling Ecosystem (Figure 4), with 1/6° of spatial resolution. The model explicits the life history and spatio-temporal dynamics of 9 species (1 macro-zooplankton group, 1 crustacean, 1 cephalopod and 6 fish species), between 1992 and 2008.

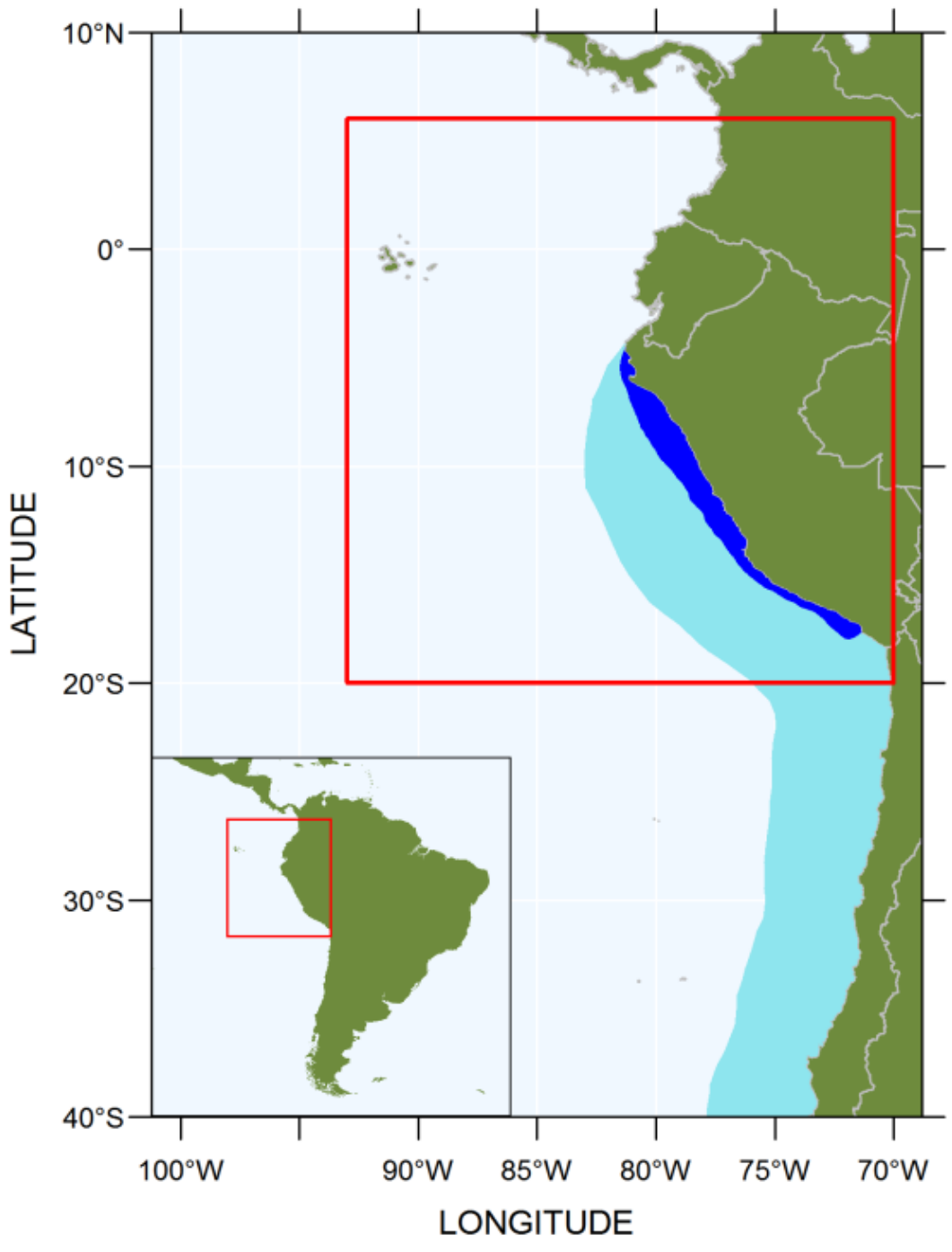

Figure 4. Map of the modeled area. The model spatial domain is limited by the red square. The light blue area shows the extension of the Humboldt Current Large Marine Ecosystem, and the dark blue area the extension of the Peruvian Upwelling Ecosystem.

The NHCE exhibits a high climatic and oceanographic variability at several scales, e.g. seasonal, interannual and decadal. The major source of interannual variability is due to the interruption of the upwelling seasonality by the El Nino Southern Oscillation ENSO (Alheit and Ñiquen 2004), which imposes direct effects on larval survival and fish recruitment success (Ñiquen and Bouchon 2004). Consequently, the fishing activity can also be highly variable, depending on the variability in the abundance and accessibility of the main fishery resources. Due to these different sources of temporal variability, it is thus important to model the temporal variability in both larval mortality and fishing mortality. This variability was modeled using time-varying parameters, which were estimated considering interannual and seasonal variability. For both processes, we estimated the mean value and annual deviations. The natural mortality (due to other sources of mortality not included in the model) and plankton accessibility to fish coefficients were also estimated.

The data used to calibrate our model were: i) biomass indices from hydro-acoustic scientific surveys (Gutierrez et. al 2000; IMARPE 2010) and ii) total reported landings for the main commercial species (IMARPE 2009, IMARPE 2010). Additionally, catch-at-length data were available for anchovy and jack mackerel and catch-at-age for hake. A summary of the data available, the time resolution of the information and the period of availability is shown in Table 1. The model was confronted to data using a likelihood approach described in Appendix A. The optimization of the likelihood function was carried out using an evolutionary algorithm developed by Oliveros-Ramos and Shin (submitted), since for stochastic models it is not possible to apply derivative-based methods (see Appendix A for more details).

Table 1.Summary of data available for the calibration of the model. Years for data availability are indicated, M (monthly) and Y (yearly) indicates the time resolution of the data.

| | **Catch-at-age** | **Catch-at-length** | **Landings** | **Acousticindex** |
|---|---|---|---|---|
| Euphausiids | | | No fishing | 2003 - 2008 (Y) |
| Anchovy (*Engraulis ringens*) | | 1992 - 2008 (M) | 1992 - 2008 (M) | 1992 - 2008 (Y) |
| Sardine (*Sardine sagax*) | | | 1992 - 2008 (M) | 1992 - 2008 (Y) |
| Jack Mackerel (*Trachurus murphyi*) | | 1992 - 2008 (Y) | 1992 - 2008 (M) | 1992 - 2008 (Y) |
| ChubMackerel (*Scomber japonicus*) | | | 1992 - 2008 (M) | 1992 - 2008 (Y) |
| Mesopelagic fish | | | No fishing | 1999 - 2008 (Y) |
| Red lobster (*Pleuroncodes monodon*) | | | No fishing | 1999 - 2008 (Y) |
| Jumbo squid (*Dosidicus gigas*) | | | 1992 - 2008 (M) | 1999 - 2008 (Y) |
| Peruvian hake (*Merluccius gayi*) | 1992-2008 (Y) | | 1992 - 2008 (M) | 1992 - 2008 (Y, trawl) |

## 2.4. Calibration experiments

For the OSMOSE model of the Northern Humboldt Current Ecosystem (NHCE), 5 types of parameters needed to be estimated (Table 3): i) larval mortality rates, ii) fishing mortality rates, iii) coefficients of plankton accessibility to fish, iv) natural mortality rates (due to other sources of mortality not explicit in OSMOSE) and, v) fishing selectivities. Additionally, we needed to estimate the immigration flux of the biomass of red lobster in the system. The calibration strategy proposed was applied to our model calibration and resulted in a pre-defined order of phases for the estimation of the parameters of the model (Table 2).

Table 2. Order of estimation of parameters in the calibration of the NHCE OSMOSE model.

| Phase | Parameters | Remarks |
|---|---|---|
| 1 | **Time-varying parameters:**<br> - **Larval mortality:** mean for all species.<br> - **Fishing mortality:** mean for all species.<br>**Non time-varying (without initial values):**<br>- **Coefficients of plankton accessibility to fish:** one for each plankton group.<br>- **Natural mortality:** all species.<br>- **Immigration:** total immigrated biomass, peak of migration flux (red lobster). | **Number of parameters estimated: 51.**<br>- Larval and fishing mortalities are assumed to vary with time.<br>- Natural mortality is assumed not to vary with time. |
| 2 | **Previous parameters**<br>**+ Time-varying parameters:**<br>- **Larval mortality:** annual deviates (all species).<br>- **Fishing mortality:** annual deviates without proxys (6 first years for squid). | **Number of parameters estimated: 208** (including previous 51).<br>- Main source of variability for the larval and fishing mortality is assumed to be interannual. |
| 3 | **Previous parameters**<br>**+ Time-varying parameters:**<br>- **Fishing mortality:** annual deviates (all remaining parameters and species). | **Number of parameters estimated: 299** (including previous 208).<br>Main source of variability for fishing mortality is assumed to be interannual. |
| 4 | **Previous parameters**<br>**+ Time-varying parameters:**<br>- **Larval mortality:** seasonal variability (anchovy).<br>**+ Non time-varying (with estimates):**<br>- **Selectivity parameters:** anchovy, hake, jack mackerel. | **Number of parameters estimated: 307** (including previous 299).<br>- Seasonal variability was only included for anchovy due to data limitations.<br>- Selectivity is assumed not to vary with time. |

We considered larval and fishing mortality as time-varying parameters, and modeled them using the functions described in appendix B (Table B.1). The immigration flux of red lobster biomass was also treated as a time-varying parameter, but is parameterized with two non time-varying parameters (total biomass immigrated in the system and the time at the peak of the migration) using a "gaussian pulse" as described in appendix B. The other parameters (natural mortality, plankton accessibility and fishing selectivities) were considered constant during the simulation and are ranked according to our evaluation of their model dependency (Figure 1).

Table 3. Models used for the time variability of larval and fishing mortalities, according to the functional forms described in Appendix B.

| *Species* | *Time-varying parameters* | | *Abundance index quality* | *Time resolution of the information* | |
|---|---|---|---|---|---|
| | *Larval mortalities* | *Fishing mortalities* | | *Catch* | *Catch at age/length* |
| **1. Abundance index, stage structured fishing information.** | | | | | |
| Anchovy | A.3 | B.2 | High | Monthly | Monthly |
| Hake | A.2 | B.2 | High | Monthly | Yearly |
| Jack mackerel | A.2 | B.2 | Low | Monthly | Yearly |
| **2. Abundance index, aggregated fishing information.** | | | | | |
| Sardine | A.2 | B.2 | High | Monthly | Not available |
| Chub mackerel | A.2 | B.2 | Medium | Monthly | Not available |
| Humboldt Squid | A.2 | B.2 | Low | Monthly | Not available |

| **3. Abundance index.** | | | | | |
|---|---|---|---|---|---|
| Red lobster | A.2 | Not required | High | - | - |
| Euphausiids | A.2 | | Low | | |
| Mesopelagics | A.2 | | Low | | |

A steady state calibration of the NHCE-OSMOSE model was performed first, corresponding to the initial conditions prevailing in 1992, which is the first year of the data time series. Red lobster was not included because it immigrates in the NHCE after 1992, during the 1997/98 El Niño event. This first calibration allowed to provide estimates of the average value of the larval and natural mortality rates for the modelled species. Fishing selectivity parameters were estimated only for the species with available age or length composition data (anchovy, hake and jack mackerel), otherwise, values from previous stock assessment were used. We used the ratio between catch and biomass, estimated directly from data, as a proxy of the variability of the fishing mortality rate. This ratio was split into seasonal and interannual deviations from the average value and used as initial values for fishing mortalities.

In order to evaluate the performance of the sequential calibration approach in the parameter estimation, we compared it to other calibration experiments:

- We first run a one-phase calibration where all parameters were estimated at once, while trying to keep constant the total number of function evaluations (i.e. the number of times we run the simulation model) to make both calibrations comparable.
- Additionally, we also run a sequential multi-phases calibration where some parameters were fixed. Concretely, we fixed the natural and fishing mortality variability to values reported in the literature. As in previous OSMOSE applications (Marzloff et al. 2009, Duboz et al. 2010), natural and fishing mortalities from the Ecopath with Ecosim model for the same system (Tam et al. 2008) were used. In this case, only larval mortalities, coefficients of fish accessibility to plankton, the immigration flux of red lobster and average fishing mortality were estimated.
- Finally, we run a calibration without including the annual deviates, estimating only the long term mean of the larval mortality and the average fishing mortality, to analyze the effect of considering time-varying parameters in the calibration results.

In all calibration experiments, we considered commercial landings data as the most reliable source of information, compared to estimates of species biomass derived from scientific surveys. In consequence, more weight was given to catch data (i.e.

less uncertainty; CV=0.05) than to the biomass indices (CV=0.15, 0.10 for anchovy), so the fit to catch data contributed the most to the total value of the likelihood function we tried to optimize during the calibration process.

## 3. Results and discussion

### 3.1 Calibration in multiple-phase

As we considered commercial landings as the most reliable source of information in the likelihood function, catch output from the NHCE-OSMOSE model should have better fit than all the information used for the calibration. The simulated landings we obtained are in good agreement with data, at monthly and yearly scale (Figures 5 and 6), with high correlations between observed and simulated values. The landings of the Humboldt squid have the poorest fitting, which can be partially explained by the lack of abundance indices and proxys for the variability in the fishing mortality for the initial years. Landings of anchovy have the highest variability, since fishing mortality has a strong seasonality (Oliveros-Ramos et al. 2010), with no fishing over several months and the setting of two quotas per year. A refined modeling of the variability of the fishing mortality for anchovy (two parameters per year, instead of one for example) may help improve fitting to the observed landings. Nevertheless, with the current calibration, the fit of the NHCE-OSMOSE model to anchovy landings can be considered as satisfactory, capturing most of the interannual variability observed for this fishery. The landings of hake are overestimated for most of the simulated period, but the trends in variability are properly reproduced, having the highest correlation between observed and simulated landings among all the species. Jack mackerel and sardine are the ones presenting the best fit, while there is a significant overestimation of the landings of chub mackerel during 1999 and 2000. The simulated landings were produced from the estimated fishing mortalities and our initial choice of the selectivity functions. The fishing mortality rates for a given length or age class are highly variable over time for all target species (Figure 7).

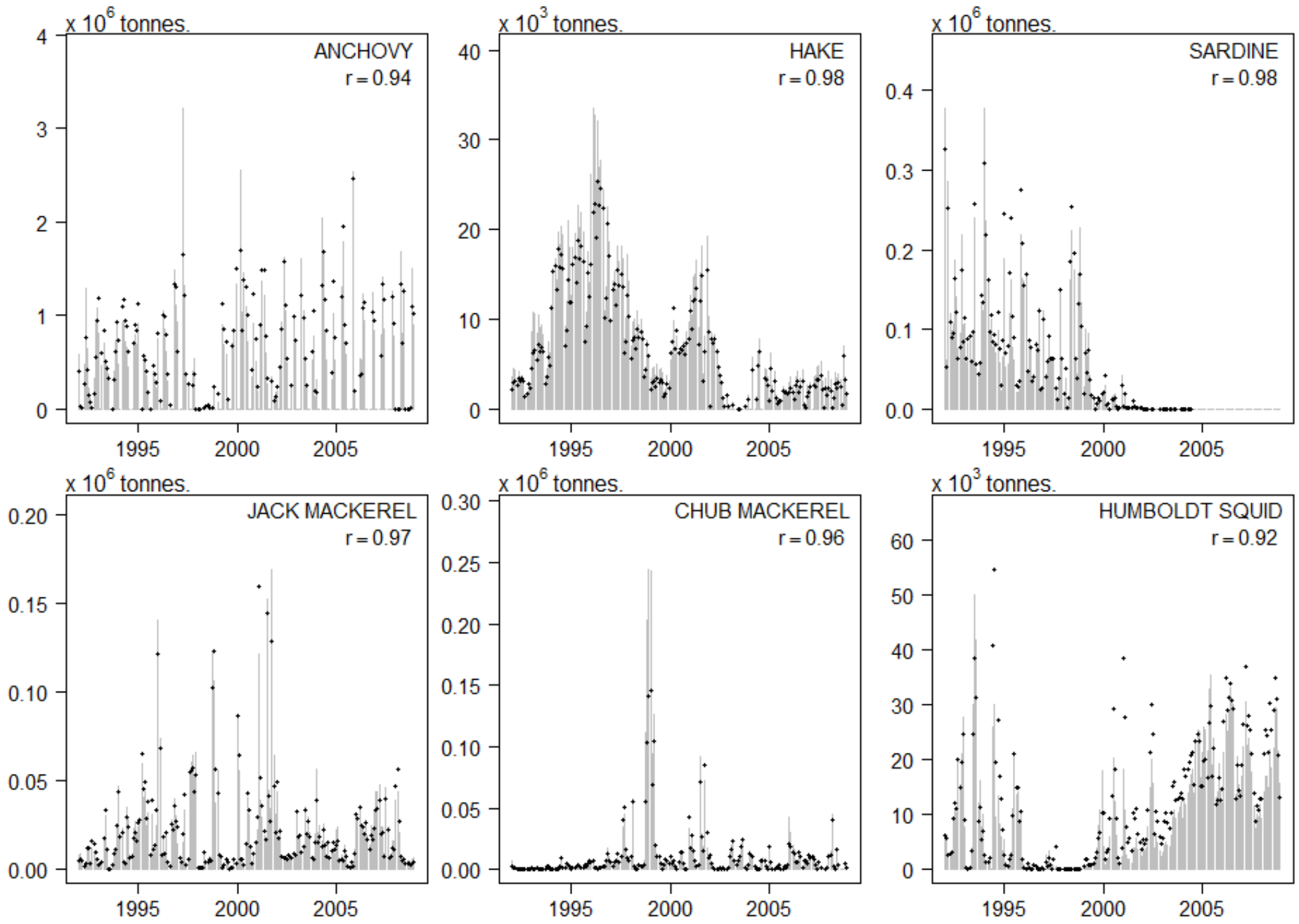

Figure 5. Fit of the NHCE-OSMOSE model to the monthly landings data for the multiple-phase reference calibration. The grey bars represent the monthly landings predicted by the model and the dots the observed landings, for each modelled species targeted by Peruvian fisheries.

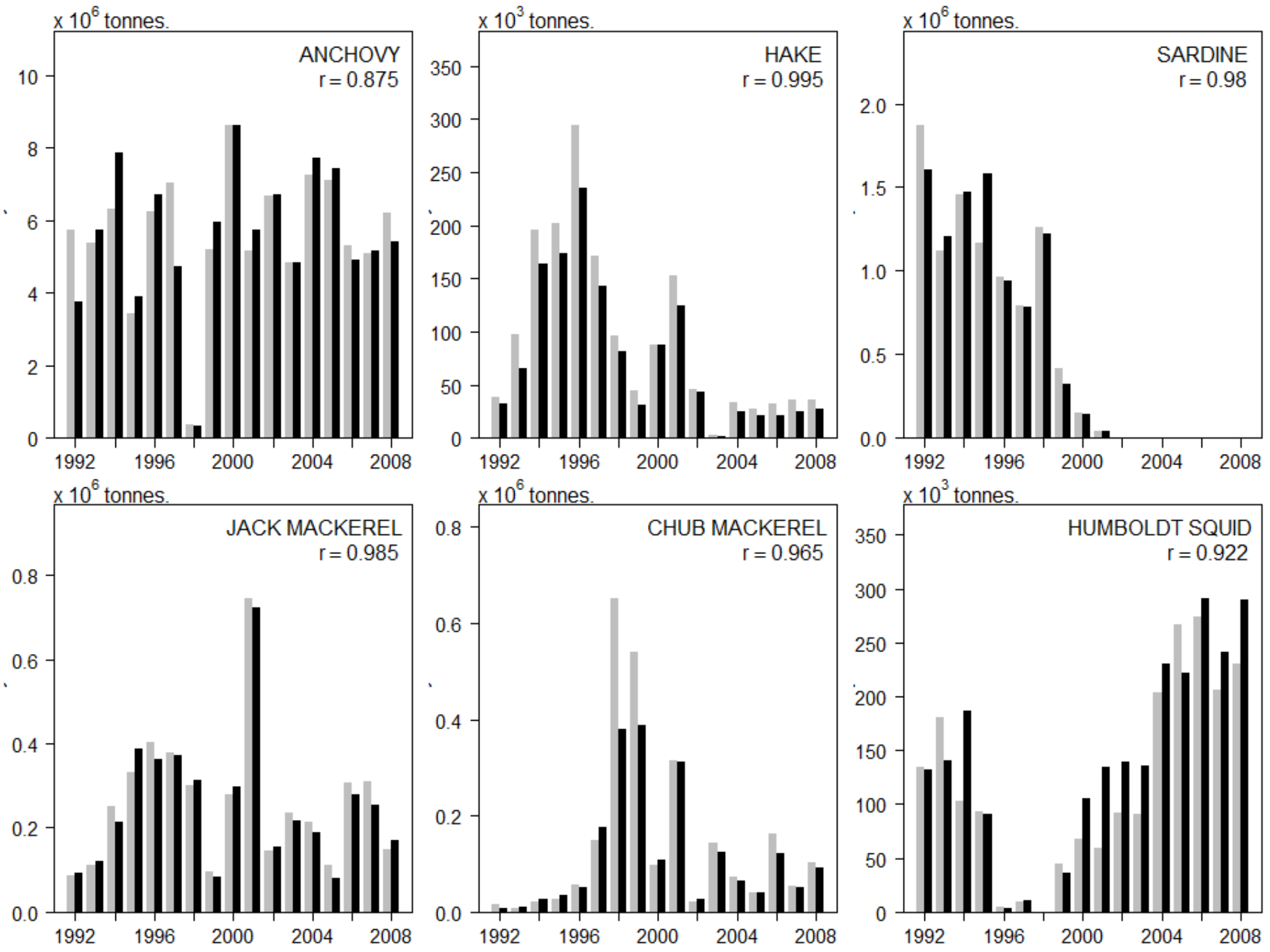

Figure 6. Fit of the NHCE-OSMOSE model to the annual landings data for the multiple-phase reference calibration. The grey and black bars represent the annual landings predicted by the model and the observed landings, respectively, for each species with active fisheries.

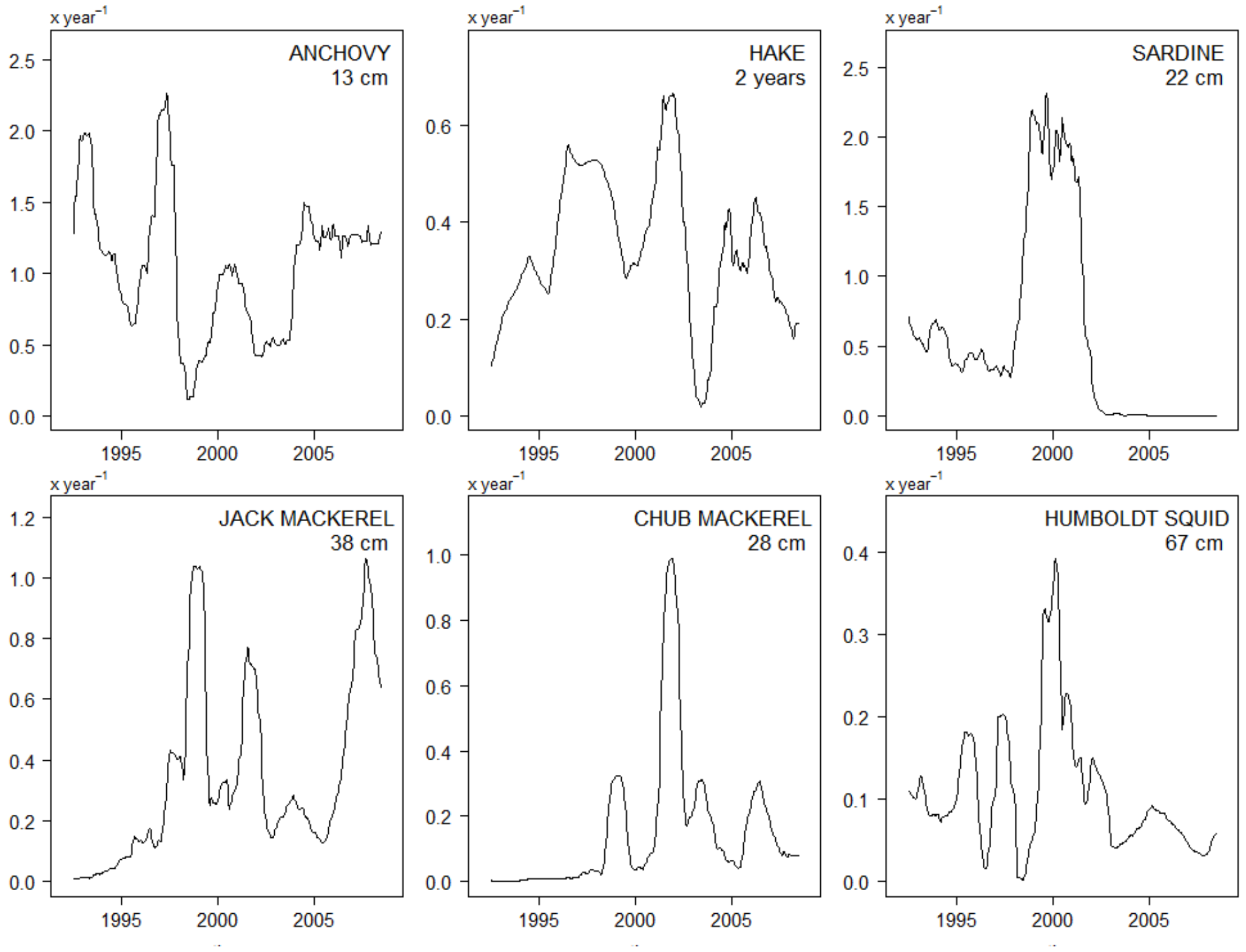

Figure 7. Estimated average fishing mortality rates ($year^{-1}$) for the most representative length or age class observed in the fishery.

For the age and length distributions of fish abundance (Figure 8), we obtained a good fit considering we assumed only one selectivity pattern over the whole studied period, while time-varying selectivity is more standard in fishery modeling (REFS). For anchovy, the model predicts more young adults (12-14cm) and less juveniles (< 12cm) in average, but in the temporal patterns of the residuals, there are both over and underestimation of fish density for the same length classes. The model does not include spatial variability in the fishing effort, which usually concentrates in the central part of Peru (IMARPE 2010). In addition, anchovy biomass spatial distribution by size can be also very variable, depending on the environmental conditions (Bertrand et al. 2004, Gutierrez et al. 2007, Swartzman et al. 2008). Including these sources of variability properly could help to explain better the variability patterns in the catch at size without introducing time variability in the selectivity. For jack mackerel, main differences are localized in a few years. The discrepancy could be due to the fact that Jack mackerel is a straddling stock which can show high variability in spatial distribution (Dioses 2013, Segura and Aliaga 2013, Ayón y Correa 2013), and we did not model the spatial distribution of fishing effort which is concentrated inside the Peruvian EEZ (Ñiquen et al. 2013, Zuzunaga 2013). In addition, the bimodal shape of the distribution of observed catch at length is related to a higher than usual proportion of juveniles in the landings in 2004 (Diaz 2013). The model is predicting more juveniles in the range of 15 to 25 cm in order to properly fit the small mode around 15cm. OSMOSE being a spatial model, these issues can possibly be handled in future by improving the spatial definition of the habitat for different classes of jack mackerel, without adding ad hoc assumptions on selectivity changes. Finally, for hake, the landings of the main age classes exploited by the trawl fishery (1 to 3 years) are well represented while landings of older classes are in general overestimated. However, these older age classes are normally not accessible to the industrial trawl fishery. This may indicate model misspecification, suggesting a simple lognormal selectivity model as used here (Appendix A) may not be appropriate.

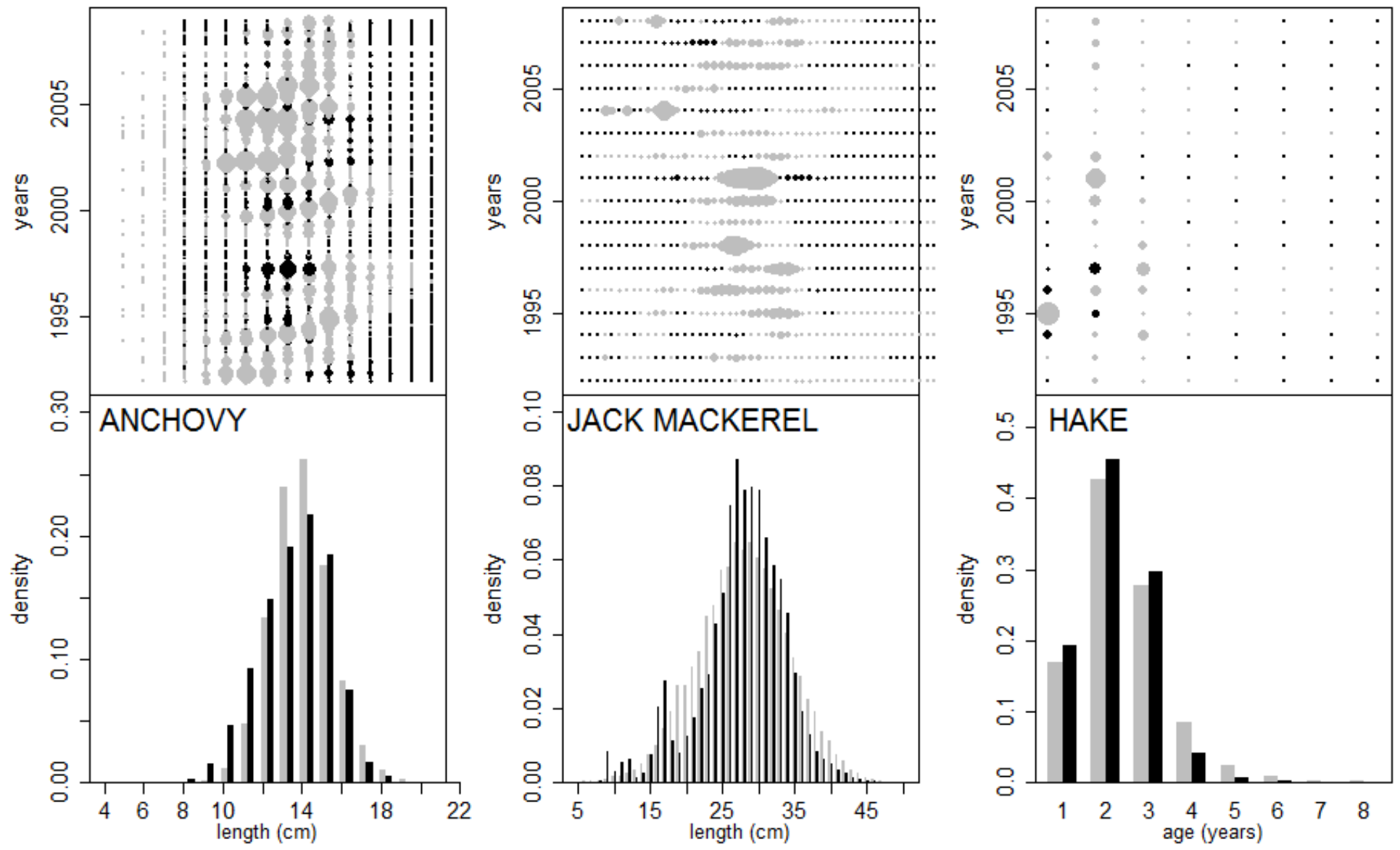

Figure 8. Fit of the NHCE-OSMOSE model to catch at length/age data for the multiple-phase reference calibration. The upper panels show the raw residuals; the size of the circle is proportional to the magnitude of the error, grey means that the simulated density is lower than observed, and black means the simulated density is higher than observed. The bottom panels show the average size or age distributions of fish abundance, grey bars represent simulated outputs and black bars observed values.

The second important source of information for the calibration was provided by the time series of surveyed biomass indices. The model captured properly the important trends in the time series of the abundance indices of all species except for mesopelagic fish and euphausiids (Figure 9). Two main problems with these two species groups could explain the poor fit: i) lack of abundance data during the first few years of the simulation and ii) lack of landings data because these two species groups are not harvested. For mesopelagic fish, the simulated biomass is within the range of observed biomass but the simulations do not reproduce the observed interannual variability pattern. For euphausiids, there is a systematic overestimation of the biomass, which combined with a low larval mortality (Figure 10) suggests a incompatibility of the model configuration with the observed biomass. Ballon et al. (2011) suggested that there might be an underestimation of euphausiids biomass directly derived from scientific surveys using traditional methods. Therefore, it is likely that the system requires more macrozooplankton biomass than what was observed from the surveys in order to sustain the observed biomass levels of other species.

The estimated larval mortalities for all species are shown in Figure 10. A more detailed analysis and validation of the estimated larval mortalities is necessary to assess the temporal signals in these parameters, since there is a potential risk that some of its variability is an artifact to fit properly the observed biomass. For species with length information, it is possible to better estimate the larval mortalities since the length information helps to disentangle possible confounding effects with other parameters like the base natural mortality (representing all other sources of mortality not included in the model) which affects all length classes uniformly.

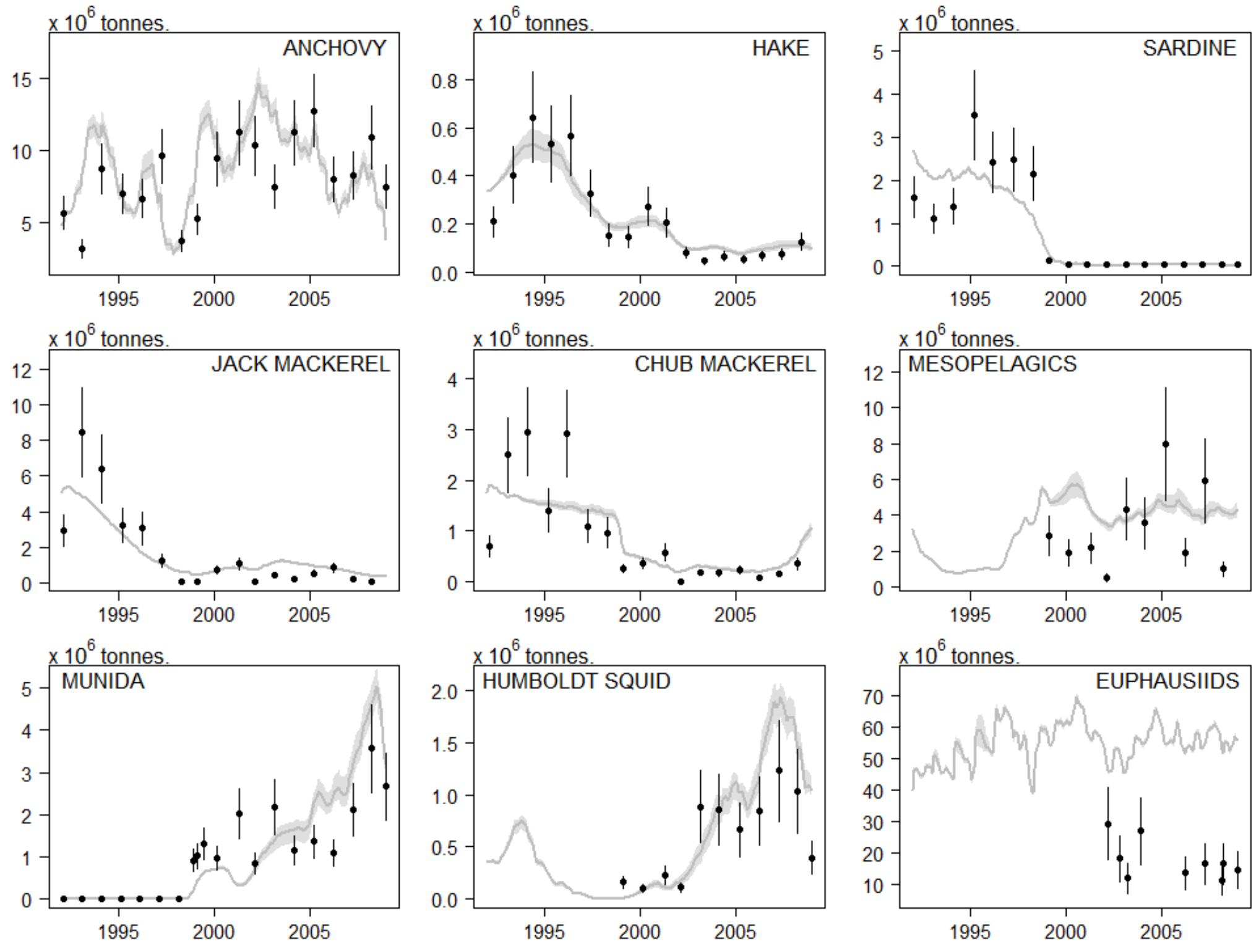

Figure 9. Fit of the NHCE-OSMOSE model to the monthly survey biomass for the reference multiple-phase calibration. The shaded area represent the 95% confidence interval for the simulated biomass, considering the model stochasticity only. The black dots and bars represent the observed value and 95% confidence intervals for the observations, given the CV assumed for the data errors in the calibration.

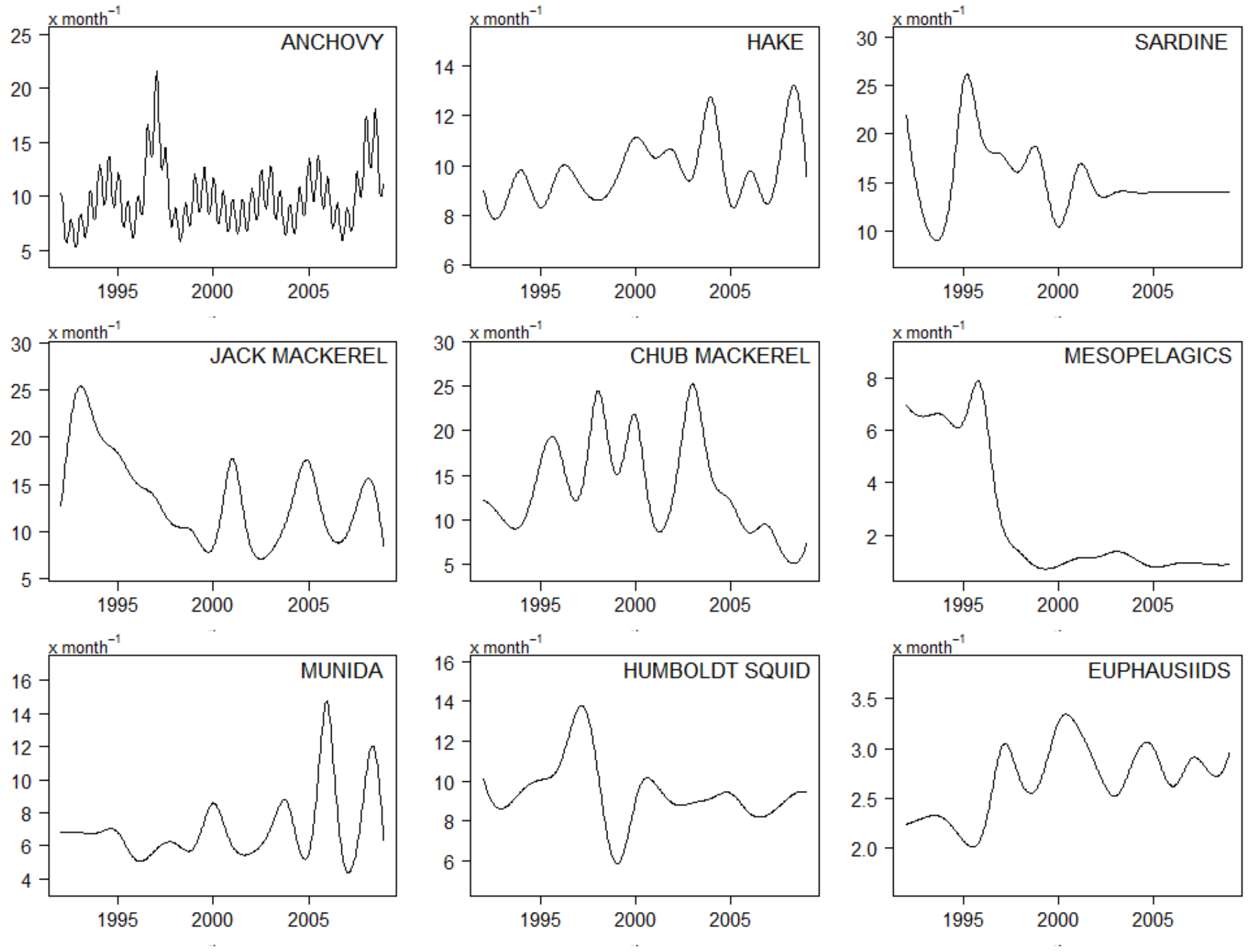

Figure 10. Larval mortality rates estimated by the reference multiple-phase calibration for all modelled species.

### 3.2 Comparison with other calibration experiments

Since the same data has been used for all the calibration experiments, the likelihood contribution of each source of information is comparable between models. In particular, the first two experiments (multiple and single phases calibrations) are directly comparable since the only difference is the strategy used for the calibration of the same model configuration and for the same number of parameters. The results show clearly that the multiple-phase calibration allowed to improve the optimisation (lower AIC and negative log-likelihood; Table 4). The calibration run with some parameters fixed from the literature, was not able to fit the landings as well as do the other calibrations, probably because the variability in the fishing mortality rates was fixed and these parameters are model-dependent. Also, there is a poorest fit to the abundance indices, which can be more related to the mis-specification of the natural mortalities. The calibration without interannual parameters is not able to fit properly the landings nor the biomass indices, and the interannual variability observed in the simulations come from the forcing effect of fish habitat distribution and from the plankton dynamics (ROMS-PISCES output forcing OSMOSE – Appendix A). Larval mortality can be strongly affected by the environment and this parameter can have an important impact, particularly for the dynamics of short lived species which depend more on the level of recruitment, like Peruvian anchovy (Oliveros-Ramos and Peña-Tercero 2011). Since OSMOSE does not include an explicit sub-model for such variability in eggs and larval survival, the estimation of the interannual variability in the larval mortality was necessary. Similar reasoning can be applied to the case of fishing mortality since the variability in this parameter can be related not only to the availability of the resource biomass but also to social and economical constraints.

Table4.Summary of the likelihood for the different calibration experiments of the NHCE-OSMOSE model.

| Calibrationexperiment | Number parameters | AIC | Negative log-likelihood | | | |
|---|---|---|---|---|---|---|
| | | | Total | Landings | BiomassIndex | Catch-at-age/length |
| Multiple-phase | 307 | 74807.8 | 37096.9 | 26958.7 | 1778.4 | 8333.2 |
| Single phase | 307 | 101030.1 | 50208.1 | 39356.8 | 2267.8 | 8543.3 |
| Fixed parameters | 207 | 142731.0 | 71158.5 | 57943.8 | 2855.9 | 10287.9 |
| Without interannual parameters | 56 | 280353.0 | 140120.5 | 128809.7 | 3319.5 | 7985.0 |

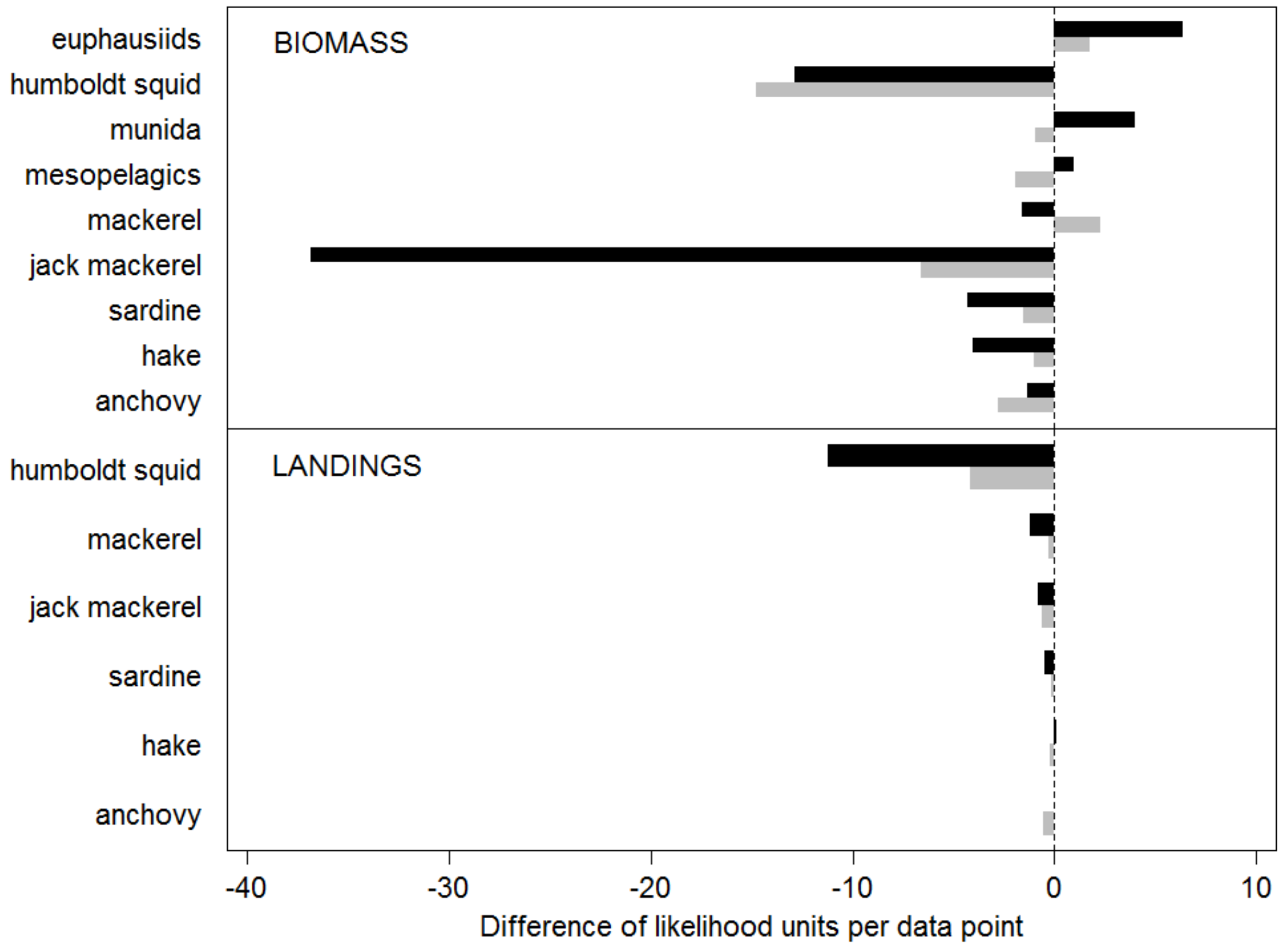

Figure 11. Comparison of the likelihood components for the different calibration experiments. The difference between the calibration experiment and the reference calibration (multiple-phase) is shown. A negative difference means that the reference calibration fitted the data better for that particular component. The one phase calibration is shown in grey and the calibration with fixed parameters in black.

Globally, the comparative calibration experiments show that the reference calibration (multiple-phase) fits better the landing data than do the other calibrations (one phase and fixed parameters), except for hake where the difference is negligible (Figure 11). The reference calibration fits better the biomass data as well for most species. However, the reference calibration clearly performs better in comparison to the other two calibrations. Considering the temporal variability of the log-normal errors for the simulated landings (Figure 12), the two calibrations with the full parameterization (one or multiple-phase) fit the data better than the calibration with fixed parameters. Hake is the exception for which the reference calibration produces a systematic overestimation of the landings. The temporal patterns of the lognormal residuals of the landings are similar for all three calibrations, which can be due to the proxys of monthly fishing mortality variability which are common for all calibration experiments. Nonetheless, the reference calibration produces consistently smaller residuals for all species except for hake. In addition, for all harvested species but hake, the total likelihood of the landings is lower in the reference multi-phases calibration experiment.

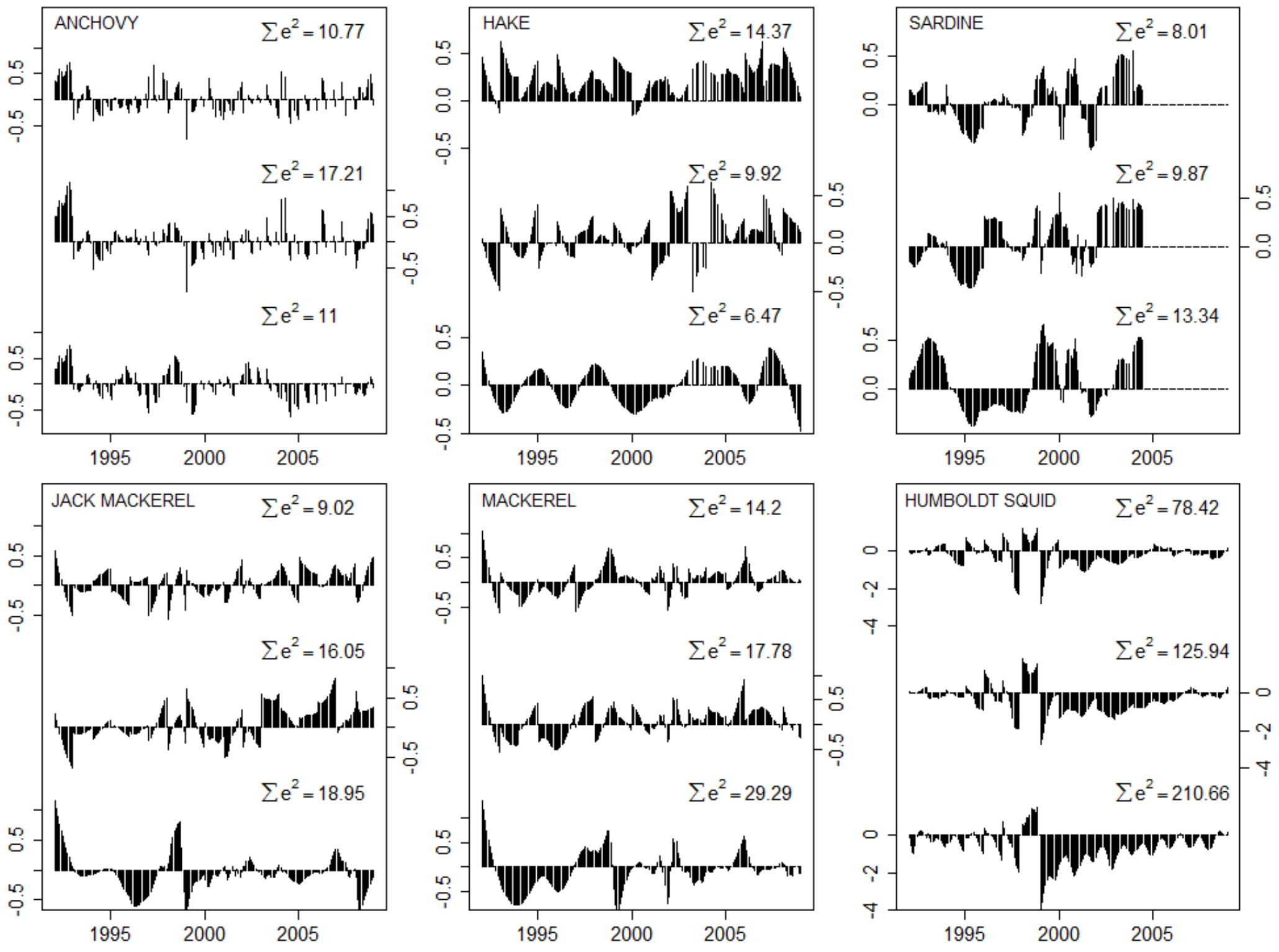

Figure 12. Comparison of the lognormal residuals of the monthly landings for the different calibration experiments. For each species, the multiple-phase (top), the single phase (center) and the fixed parameters (bottom) calibration results are shown.

The calibration experiments were also compared with regard to the predicted species biomass (Figure 13). The simulated trends are very similar across the calibrations for some species (i.e. hake, sardine and chub mackerel) while more discrepancies are present for other species. In particular, the reference calibration captured better the interannual variability for anchovy and Humboldt squid. All the calibrated models failed to reproduce the dynamics of the biomass of mesopelagic fish and euphausiids. However, the behavior of the simulations for these species shared a common pattern: steady biomass for both species, overestimation of the euphausiids biomass and average biomass of mesopelagic fish in the range of observed biomass.

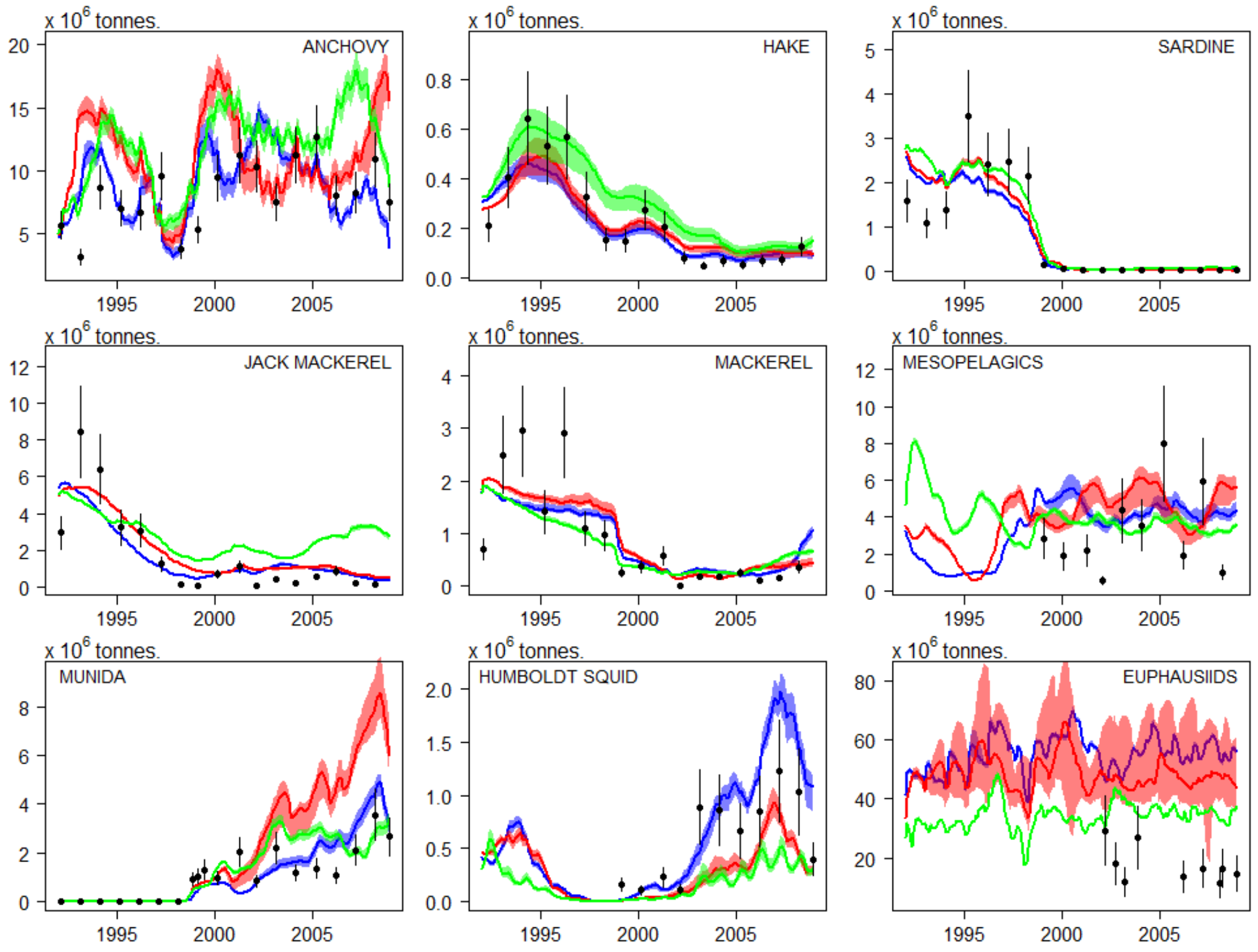

Figure 13. Comparison of the fit of the NHCE-OSMOSE model to surveyed biomass in the different calibration experiments. The shaded area represent the 95% confidence interval for the simulated biomass and considering the model stochasticity only. The black dots and bars represent the observed value and 95% confidence intervals for the observations, given the CV assumed for the data in the calibration. The calibration fixing parameters from other models estimates (green) is compared to the one phase calibration (red). The reference calibration in multiple-phase (blue) is also shown.

For all our experiments, we have used the standard information used in fishery models (landings, abundance indices, catch at age or length). Since ecosystem models can provide more outputs to be confronted to data (e.g. diets, size spectrum, and other community indicators), the availability and use of these additional information could help selecting the more appropriate parameterization among different alternatives. On the other hand, if this type of data is not available, this could direct new objectives in data collection (Rose 2012) which can lead to important and necessary improvements of ecosystem models in general.

## 4. Conclusions and perspectives

Using a dedicated global search optimization method (Oliveros-Ramos and Shin, in prep.), we proposed a sequential multi-phases calibration approach which allowed to improve significantly the estimation of the model parameters and to lead to a better agreement between the model and the data. Our main objective was to provide guidelines to improve the calibration of ecosystem models. We focused on model dependency and time variability to categorize the parameters since these are two criteria which are usually considered to reduce the number of parameters to estimate. This preliminary parameters' classification can lead to fixing parameter values from other models or species/ecosystems or to ignoring time variability in the parameters (Lehuta et al 2013). However, in the present study, we have not considered other useful criteria such as sensitivity analysis, which has been used to reduce the number of parameters to be estimated (Megrey et al. 2007, Dueri et al. 2012, Lehuta et al 2013). Additionally, a successful calibration does not mean that a model is reliable (Gaume et al. 1998), and a proper validation is always required, eventually providing information to improve the model and to revise the calibration (Walter and Pronzato 1997, Jorgensen and Bendoricchio 2001).

**Acknowledgements**

The authors were partly funded through the French project EMIBIOS (FRB, contract no. APP-SCEN-2010-II). ROR was supported by an individual doctoral research grant (BSTD) from the "Support and training of scientific communities of the South" Department of IRD, managed by Egide. We thank the staff of IMARPE for collecting and processing all the data used in this paper, and Coleen Moloney and Arnaud Bertrand for their support and advice during the development of this research. All the calibration experiments were performed with the IFREMER-CAPARMOR HPC 600 facilities. This work is a contribution from the Cooperation agreement between the Instituto del Mar del Peru (IMARPE) and the Institut de Recherche pour le Developpement (IRD), through the LMI DISCOH. The editors of the special volume dedicated to Bernard Megrey are warmly thanked for their initiative.

**References.**

Alheit J., Ñiquen M., 2004. Regime shifts in the Humboldt Current ecosystem. Progress in Oceanography 60:201–222.

Ayón P. and Correa J., 2013. Spatial and temporal variability of Jack mackerel *Trachurus murphyi* larvae in Peru between 1966 –2010. Rev. peru. biol. 20(1):083- 086.

Bäck T., Schewefel H.-P., 1993. An Overview of Evolutionary Algorithms for parameter optimization. Evolutionary Computation 1:1-23.

Ballon M., Bertrand A., Lebourges D. A., Gutierrez M., Ayon P., Grados D., Gerlotto F., 2011. Is there enough zooplankton to feed forage fish populations off Peru? An acoustic (positive) answer. Progress in Oceanography 91 (4):360-381.

Bartell S.M., 2003. Effective use of ecological modeling in management: The toolkit concept. In Dale V (editor).Ecological modeling for resource management, Springer Verlag.

Bakun A. and Parrish R., 1982. Turbulence, transport, and fish in California and Peru currents. Calcofi Rep. Vol. xxiii. 99-112.

Bertrand A., Segura M., Gutiérrez M., Vásquez L., 2004. From small-scale habitat loopholes to decadal cycles: a habitat-based hypothesis explaining fluctuation in pelagic fish populations off Peru. Fish and Fisheries 5:296–316.

Bolker B.M., 2008. Ecological models and Data in R. Princeton University Press. 408pp.

Bolker B.M., Gardner B., Maunder M., Berg C.W. , Brooks M., Comita L., Crone E., Cubaynes S., Davies T., de Valpine P., Ford J., Gimenez O., Kéry M., Kim E.J., Lennert-Cody C., Magnusson A., Martell S., Nash J., Nielsen A., Regetz J., Skaug H., Zipkin E., 2013. Strategies for fitting nonlinear ecological models in R, AD Model Builder, and BUGS. Methods in Ecology and Evolution 4: 501–512.

Bundy A., 2005. Structure and functioning of the eastern Scotian Shelf ecosystem before and after the collapse of ground fish stocks in the early 1990s. Canadian Journal Of Fisheries And Aquatic Sciences 62:1453-1473.

Diaz E., 2013. Estimation of growth parameters of Jack mackerel Trachurus murphyi caught in Peru from length frequency analysis. Rev. peru. biol. 20(1):061- 066.

Dioses T., 2013. Abundance and distribution patterns of Jack mackerel Trachurus murphyi in Peru. Rev. peru. biol. 20(1):067- 074.

Duboz R., Versmisse D., Travers M., Ramat E., Shin Y.-J., 2010. Application of an evolutionary algorithm to the inverse parameter estimation of an individual-based model. Ecological Modelling 221(5):840-849.

Dueri S., Faugeras B., Maury O., 2012. Modelling the skipjack tuna dynamics in the Indian Ocean with APECOSM-E – Part 2: Parameter estimation and sensitivity analysis. Ecological Modelling 245:55-64.

Echevin V., Goubanova K., Dewitte B., Belmadani A., 2012. Sensitivity of the Humboldt Current system to global warming: a downscaling experiment of the IPSL-CM4 model, Climate Dynamics. doi:10.1007/s00382-011-1085-2.

Fournier D.A., Skaug H.J., Ancheta J., Ianelli J., Magnusson A., Maunder M.N., Nielsen A., Sibert J., 2012. AD Model Builder: using automatic differentiation for statistical inference of highly parameterized complex nonlinear models. Optimization Methods and Software, 27:2, 233-249, DOI: 10.1080/10556788.2011.597854.

Fournier D.A., 2013. An introduction to AD Model Builder for use in Nonlinear Modeling and Statistics. ADMB Foundation, Honolulu. 212pp.

Friska M.G., Miller T.J., Latour R.J., Martell S.J.D., 2011. Assessing biomass gains from marsh restoration in Delaware Bay using Ecopath with Ecosim. Ecological Modelling 222:190–200.

Gaume E., Villeneuve J.-P., Desbordes M., 1998. Uncertainty assessment and analysis of the calibrated parameter values of an urban storm water quality model. Journal of Hydrology 210: 38–50.

Guénette S., Christensen V., Pauly D., 2008. Trophic modelling of the Peruvian upwelling ecosystem: Towards reconciliation of multiple datasets. Progress in Oceanography 79: 326–335.

Gutiérrez M. et al., 2000. Estimados de biomasa hidroacústica de los cuatro principales recursos pelágicos en el mar peruano durante 1983 -2000. Bol. Inst. Mar Perú. Vol. 19, n°1-2, pp. 136-156.

Gutiérrez M., Swartzman G., Bertrand A., Bertrand S., 2007. Anchovy (*Engraulis ringens*) and sardine (*Sardinops sagax*) spatial dynamics and aggregation patterns in the Humboldt Current ecosystem, Peru, from 1983–2003. Fish. Oceanogr. 16:2, 155–168.

IMARPE, 2009. Informe sobre la tercera reunión de expertos en dinámica de evaluación de la merluza peruana. Bol. Inst. Mar. Perú – Callao. Bol. Inst. Mar Perú 24(1-2).

IMARPE, 2010. Informe sobre la quinta reunión de expertos en dinámica de población de la anchoveta peruana. Bol. Inst. Mar. Perú – Callao. Bol. Inst. Mar Perú 23(1-2).

Jones G., 1998. Genetic and evolutionary algorithms. Encyclopedia of Computational Chemistry. John Wiley and Sons.

Jorgensen S.E. and Bendoricchio G., 2001. Fundamentals of Ecological Modelling. Third Edition. Elsevier. 530pp.

Lehuta S., Mahévas S., Petitgas, P., Pelletier D., 2010. Combining sensitivity and uncertainty analysis to evaluate the impact of management measures with ISIS-Fish: marine protected areas for the Bay of Biscay anchovy (Engraulis encrasicolus) fishery. ICES J. Mar. Sci. 67, 1063–1075.

Lehuta S., Petitgas P., Mahévas S., Huret M., Vermard Y., Uriarte A., Record N.R., 2013. Selection and validation of a complex fishery model using an uncertainty hierarchy. Fisheries Research 143:57- 66.

Mackinson S., and Daskalov G., 2007. An ecosystem model of the North Sea for use in research supporting the ecosystem approach to fisheries management: description and parameterisation [online]. (CEFAS, Lowestoft.) Available from www.cefas.co.uk/publications/techrep/tech142.pdf.

Marzloff M., Shin Y.-J., Tam J., Travers M., Bertrand A., 2009. Trophic structure of the Peruvian marine ecosystem in 2000–2006: Insights on the effects of management scenarios for the hake fishery using the IBM trophic model Osmose. Journal of Marine Systems 75: 290-304.

Maunder M.N. and Deriso R.B., 2003. Estimation of recruitment in catch-at-age models. Can. J. Fish. Aquat. Sci.60: 1204–1216.

Megrey B.A., 1989. A Review and Comparison of Age-Structured Stock Assessment Models from Theoretical and Applied Points of View. NWAFC Processed Report 88-21. 124pp.

Megrey B.A., Rose K.A, Klumb R.A, Hay D.E, Werner F.E, Eslinger D.L, Smith S.L., 2007. A bioenergetics-based population dynamics model of Pacific herring (Clupea harengus pallasi) coupled to a lower trophic level nutrient–phytoplankton–zooplankton model: Description, calibration, and sensitivity analysis. Ecological Modelling 202:144–164.

Methot R.D. Jr. and Wetzel C.R., 2013. Stock synthesis: A biological and statistical framework for fish stock assessment and fishery management. Fisheries Research 142: 86–99.

Nash J.C. and Walker-Smith M., 1987. Nonlinear Parameter Estimation: an Integrated System in BASIC. Marcel Dekker, New York. 493pp.

Ñiquen M., Bouchon M., Ulloa D., Medina A., 2013. Analysis of the Jack mackerel Trachurus murphyi fishery in Peru. Rev. peru. biol. 20(1):097-106.

Ñiquen M, Bouchon M., 2004. Impact of El Niño event on pelagic fisheries in Peruvian waters. Deep-Sea Research II 51:563-574.

Oliveros-Ramos R., Guevara-Carrasco R., Simmonds J., Cirske J., Gerlotto F., Peña-Tercero C., Castillo R., Tam J., 2010. Modelo de evaluación integrada del stock norte-centro de la anchoveta peruana Engraulis ringens. BolInst mar Perú 25(1-2):49-55.

Oliveros-Ramos R., Peña-Tercero C., 2011. Modeling and analysis of the recruitment of Peruvian anchovy (Engraulis ringens) between 1961 and 2009. Ciencias Marinas 37(4B):659-674.

Oliveros-Ramos R. and Shin Y.-J. Unpublished results. calibraR: an R package for the calibration of individual based models. (Submitted to Methods in Ecology and Evolution).

Oliveros-Ramos et al. Unpublished results. An end-to-end model ROMS-PISCES-OSMOSE of the northern Humboldt Current Ecosystem. In preparation.

Oliveros-Ramos et al. Unpublished results. Pattern oriented validation of habitat distribution models: application to the potential habitat of main small pelagics in the Humboldt Current Ecosystem. In preparation.

Rose K., 2012. End-to-end models for marine ecosystems: Are we on the precipice of a significant advance or just putting lipstick on a pig? Scientia Marina 76:195-201.

Ruiz D.J, Wolff M., 2011. The Bolivar Channel Ecosystem of the Galapagos Marine Reserve: Energy flow structure and role of keystone groups. Journal of Sea Research 66 (2011) 123–134.

Schnute J.T., 1994. A general framework for developing sequential fisheries models. Can. J. Fish. Aquat. Sci. 51:1676-1688.

Segura M. and Aliaga A., 2013. Acoustic biomass and distribution of Jack mackerel Trachurus murphyi in Peru. Rev. peru. biol. 20(1):087- 096.

Shannon L.J., Moloney C.L., Jarre A., Field J.G., 2003. Trophic flows in the southern Benguela during the 1980s and 1990s. Journal of Marine Systems 39:83 – 116.

Shin Y.-J., Cury P., 2001. Exploring fish community dynamics through size-dependent trophic interactions using a spatialized individual-based model. Aquatic Living Resources, 14(2): 65-80.

Shin Y.-J., Cury P., 2004. Using an individual-based model of fish assemblages to study the response of size spectra to changes in fishing. Canadian Journal of Fisheries and Aquatic Sciences, 61: 414-431.

Shin Y.-J., Rochet M.-J., Jennings S., Field J., Gislason H., 2005. Using size-based indicators to evaluate the ecosystem effects of fishing. ICES Journal of marine Science 62(3): 394-396.

Sonnenborg T.O., Christensen B.S.B., Nyegaard P., Henriksen H.J., Refsgaard C., 2003. Transient modeling of regional groundwater flow using parameter estimates from steady-state automatic calibration. J. Hydrol. (Amsterdam) 273:188–204.

Swartzman G., Bertrand A., Gutiérrez M., Bertrand S., Vasquez L., 2008. The relationship of anchovy and sardine to water masses in the Peruvian Humboldt Current System from 1983 to 2005. Progress in Oceanography 79 (2008) 228–237.

Tam J., Taylor M.H., Blaskovic V., Espinoza P., Ballón M., Díaz E., Wosnitza-Mendo C., Argüelles J., Purca S., Ayón P. et al., 2008. Trophic modeling of the Northern Humboldt Current Ecosystem. Part I: Comparing trophic linkages under La Niña and El Niño conditions. Progress In Oceanography 79(2-4):352-365.

Travers M., Shin, Y.-J., Jennings, S., Machu, E., Huggett, J.A., Field, J.G., Cury, P.M., 2009. Two-way coupling versus one-way forncing of plankton and fish models to predict ecosystem changes in the Benguela. Ecological modelling 220: 3089-3099.

Travers-Trolet M., Y.-J. Shin & J.G. Field., 2013. An end-to-end coupled model ROMS-N2P2Z2D2-OSMOSE of the southern Benguela foodweb: parameterisation, calibration and pattern-oriented validation, African Journal of Marine Science, 36:1, 11-29, DOI:10.2989/1814232X.2014.883326

Walter E. and Pronzato L., 1997. Identification of parametric models from Experimental data. Springer Masson. 413pp.

Whitley R., Taylor D., Macinnis-Ng C., Zeppel M., Yunusa I., O'Grady A., Froend R., Medlyn B. and Eamus D., 2013. Developing an empirical model of canopy waterflux describing the common response of transpiration to solar radiation and VPD across five contrasting woodlands and forests. Hydrol. Process. 27: 1133–1146.

Zuzunaga J., 2013. Conservation and fishery management regulations of Jack mackerel Trachurus murphyi in Peru. Rev. peru. biol. 20(1):107 - 113.

## Appendix A. Description of the OSMOSE model for the Northern Humboldt Current Ecosystem.

For the NHCE OSMOSE model, we considered 13 species (Table A.1), 9 being explicitly modeled in OSMOSE (1 macrozooplankton, 1 crustacean, 1 cephalopod and 6 fish species) and 4 plankton groups being represented in the ROMS-PISCES model. Total plankton biomass and average distribution from ROMS-PISCES model during the study period is shown in Figure A.1. Using Generalized Additive Models, we built maps for the spatial distribution of the species explicitly modeled in OSMOSE (Oliveros-Ramos et al. in prep. b). Providing the probability of occurrence of a species given some environmental predictors (temperature, salinity, chlorophyll-a, oxygen and bathymetry), annual maps were produced with seasonal time resolution for all species (4 maps per year) except euphausiids, for which monthly resolution (12 maps per year) was used. The average spatial distributions over the modeled period for each species are shown in Figure A.2.

Table A.1. Species or functional groups considered in the NHCE OSMOSE model. The main representative species of the functional groups are marked with an asterisk.

| Group | Species or functional groups | Scientific name | Model |
|---|---|---|---|
| Phytoplankton | Nanophytoplankton | | ROMS-PISCES |
| | Diatoms | | ROMS-PISCES |
| Zooplankton | Microzooplankton | | ROMS-PISCES |
| | Mesozooplankton | | ROMS-PISCES |
| | Euphausiids | *Euphausia mucronata** | OSMOSE |
| Small pelagics | Anchovy | *Engraulis ringens* | OSMOSE |
| | Sardine | *Sardinops sagax* | OSMOSE |
| Medium pelagics | Jack Mackerel | *Trachurus murphyi* | OSMOSE |
| | Chub Mackerel | *Scomber japonicus* | OSMOSE |
| Other pelagics | Mesopelagics | *Vinciguerria sp.** | OSMOSE |
| | Red lobster | *Pleuroncodes monodon* | OSMOSE |
| | Humboldt squid | *Dosidicus gigas* | OSMOSE |
| Demersal | Peruvian hake | *Merluccius gayi peruanus* | OSMOSE |

To compare the model biomass with the estimated by scientific surveys, we estimate a catchability coefficient $q$ from the average ratio between the distribution area and the area covered by the scientific surveys. This value was close to 1 for five species with a more coastal distribution and a good coverage of the surveys (anchovy, sardine, chub mackerel, red lobster and hake), while lower than 1 for four species (jack mackerel, mesopelagics, euphausiids and Humboldt squid).

We considered a constant selectivity over the whole model period, but used different models (logistic, normal and lognormal) for each species. A logistic selectivity was used for sardine, chub mackerel and Humboldt squid; a normal selectivity for anchovy and jack mackerel; and a log-normal selectivity for hake. All selectivities were length-based but for hake we used an age-based selectivity.

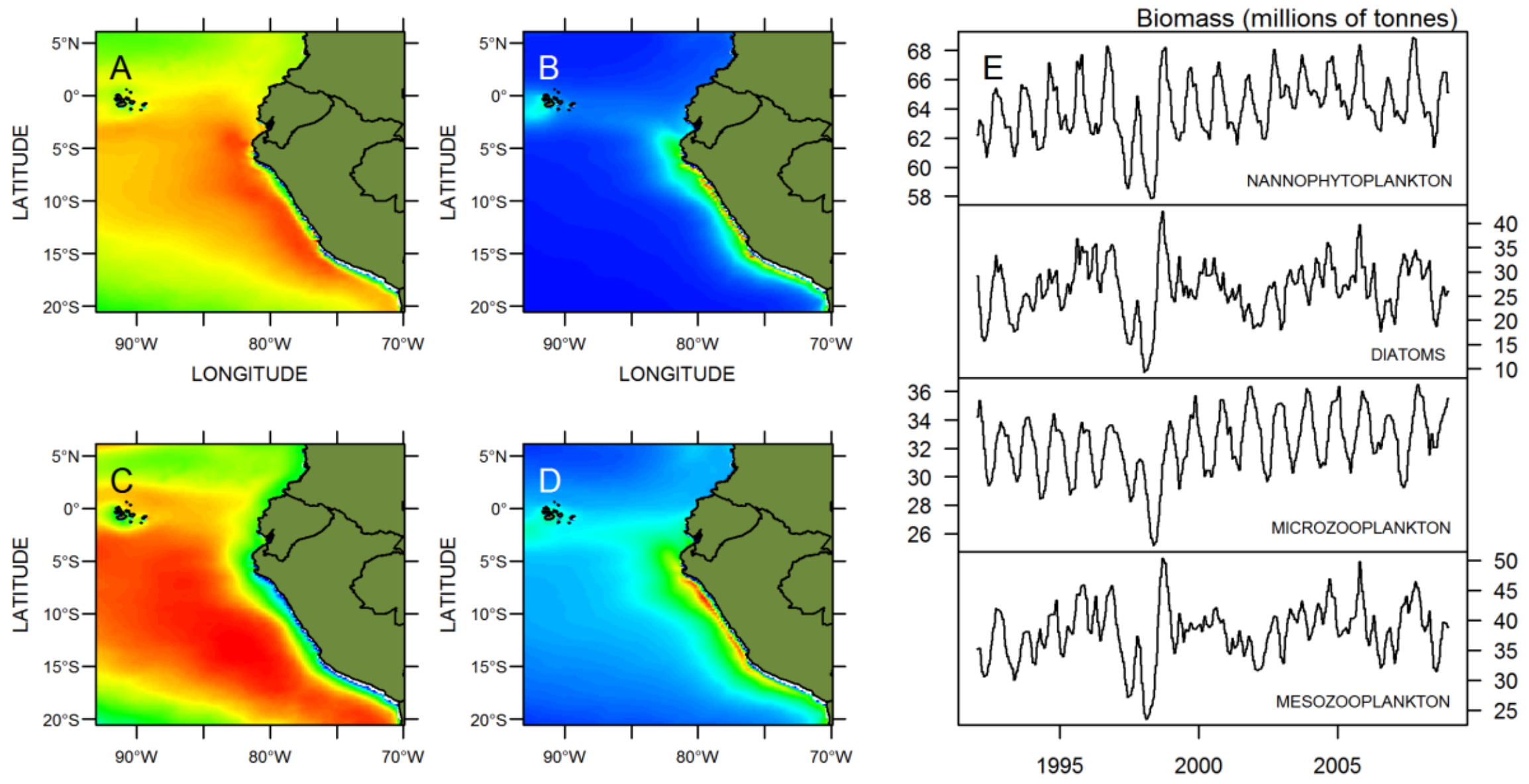

Figure A.1. Summary of the LTL biomass simulated by ROMS-PISCES model and forcing OSMOSE. Average spatial distribution for nanophytoplankton (A), diatoms (B), microzooplankton (C) and mesozooplankton (D) (red is high, blue is low, following the light visible spectrum). Simulated temporal dynamics of the total biomass (millions of tonnes) of the four plankton groups (E) is also shown.

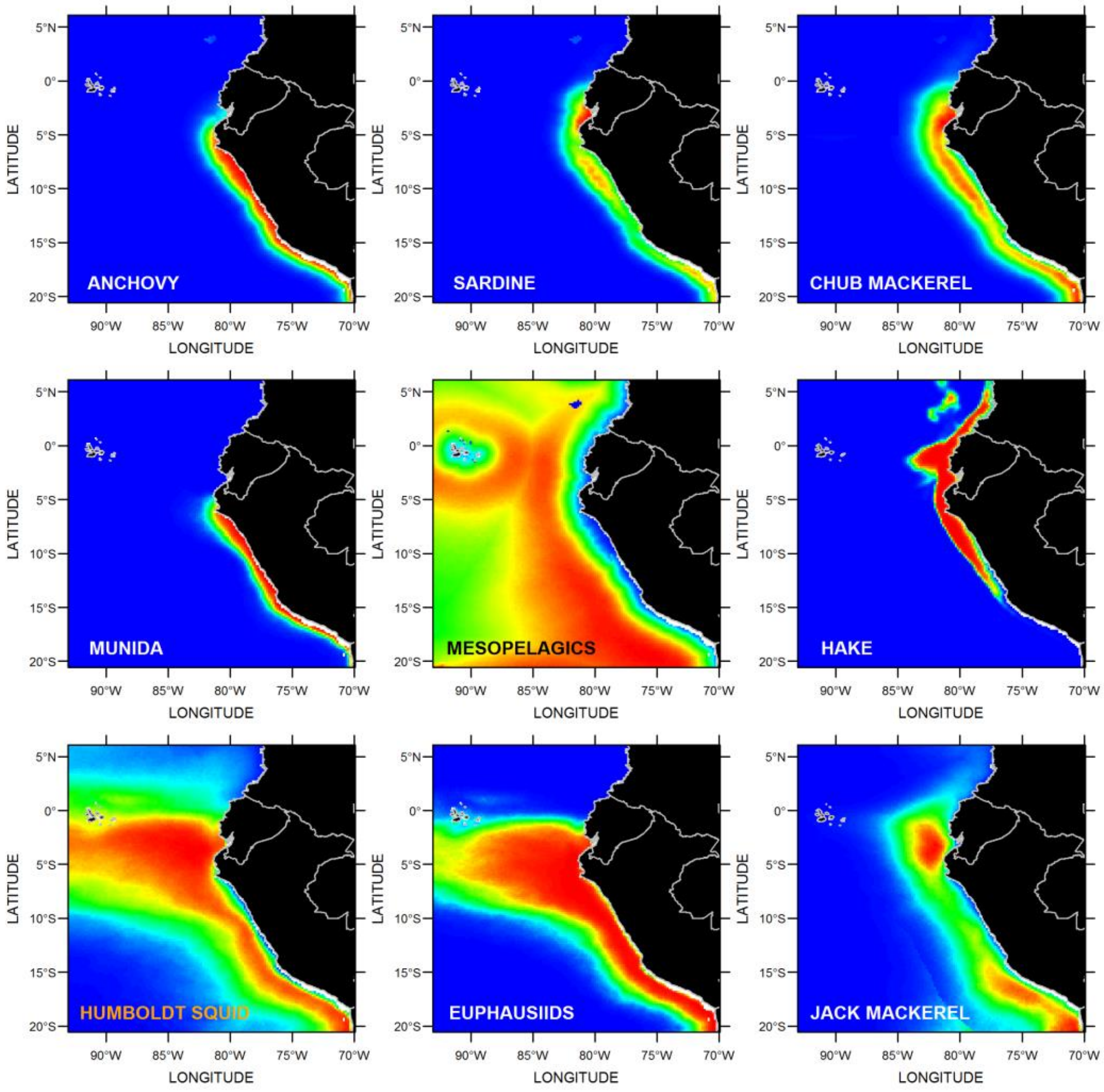

Figure A.2. Average probability distribution maps for OSMOSE species as predicted by generalized additive models (Oliveros-Ramos et al. in prep.). Probability distributions are constructed from the GAM outputs (red is high, blue is low, following the light visible spectrum).

The objective function for the calibration was a penalized negative log-likelihood function. For the likelihoods, we considered three main components: i) the errors in the biomass indices (e.g. acoustic, trawl), ii) the errors in the landings and iii) the errors in the proportions of catch-at-length or catch-at-age. A log-normal distribution was assumed for the biomass indices and landings errors, while for the age and length composition data the likelihood proposed by Maunder and Deriso (2003) was used. We also added penalties to constrain the variability in the time-varying parameters, in order to avoid overfitting. A full description of the components of the objective function is providedin Table A.2.

Table A.2. Components of the objective function.

| Likelihood/Penalty component | Equations of the likelihoods | Remarks |
|---|---|---|
| **Likelihoods** | | |
| Biomass Index | $$L_1 = \sum_s \lambda_{s,1} \sum_t \log\left(\frac{q_s B_s(t) + 0.01}{\hat{I}_s(t) + 0.01}\right)^2$$ | $\lambda_{s,1} = 22.2$ for all species s but anchovy $\lambda_{anchovy,1} = 50$ |
| Monthly Landings | $$L_2 = \sum_s \lambda_{s,2} \sum_t \log\left(\frac{Y_s(t) + 0.01}{\hat{Y}_s(t) + 0.01}\right)^2$$ | $\lambda_{s,2} = 200$ for all exploited species s |
| Catch-at-length | $$L_3 = \sum_s T_{s,l} \sum_{y=1}^{T} \sum_l -\ln\left[\exp\left(\frac{-(P_s^l(y) - \hat{P}_s^l(y))^2}{2\sigma_s^2}\right) + 10^{-3}\right]$$ | $T_{anchovy,l} = 5$ $T_{jack_mackerel,l} = 10$ |
| Catch-at-age | $$L_4 = \sum_s T_{s,a} \sum_{y=1}^{T} \sum_a -\ln\left[\exp\left(\frac{-(P_s^a(y) - \hat{P}_s^a(y))^2}{2\sigma_s^2}\right) + 10^{-3}\right]$$ | $T_{hake,a} = 10$ |
| **Penalties** | | |
| Larval mortality annual deviates | $$P_1 = \sum_s p_{s,3} \sum_{y=1}^{T} \Lambda_y$$ | $p_{s,3} = 2$ for all species but $p_{anchovy,3} = 1$, $p_{squid,3} = 8$, $p_{euphausiid,3} = 4$. |
| Natural mortality monthly deviations from proxy | $$P_2 = \sum_s p_{s,4} \sum_{y=1}^{T} m(t)$$ | $p_{s,4} = 0.5$ for all species |
| **Objective function** | | |
| | $$L = \sum_{i=1}^{5} L_i + \sum_{i=1}^{2} P_i$$ | |

The optimization problem related to minimizing the negative log-likelihood ***L*** was solved using an evolutionary algorithm developed by Oliveros-Ramos and Shin (submitted), since for stochastic models it is not possible to apply derivative-based methods (e.g. gradient descent or quasi-Newton methods). Evolutionary algorithms (EAs), which are meta-heuristic optimization methods inspired by Darwin's theory of evolution (Jones 1998), have shown their capability to yield good approximate solutions in cases of complicated multimodal, discontinuous, non-differentiable, or even noisy or moving response surfaces of optimization problems (Bäck and Schewefel 1993). They prove to be useful alternatives for the calibration of stochastic and complex non-linear models. In this EA, different parameter combinations are tested as possible solutions to minimize the ***L*** function. At each generation (i.e iteration of the optimization process), the algorithm calculates an "optimal parent" which results from the recombination of the parameter sets which provide the best solution for each objective (e.g. likelihood for biomass, yield, age/length structure). The optimal parent is then used to produce a new set of parameter combinations. To calculate this optimal parent, potential solutions are weighted according to the variability of each parameter across generations, using the coefficient of variation to take into account differences in the order of magnitude between different parameters.

## Appendix B. Models used for time-varying parameters.

We considered three type of models to represent the variability of time-varying parameters (A, B and C-type models).

The A-type models assume continuous changes in the parameter value, with a smooth interannual variability and a periodic seasonality linked to environmental drivers. The B-type models assume discrete changes for the interannual variability (e.g. driven by annual changes in management measures), and with a seasonal pattern potentially very variable between years (Table B.1). These models allow us to define nested models, for example, for A and B type models, by setting to zero the monthly and yearly effects, we reduce the parameterization to A.0 or B.0, respectively.

Finally, the C-type models assume the parameters taken non-zero values only for a shorter period of time, in this case following a Gaussian.

Table B.1 Different models used for temporal variability in parameters. For each model, X(t) is the value of the parameter at time $t$, $\bar{x}$ is the mean value of the parameter, m(t) and y(t) the month and year, respectively, at time t. The equations and the number of parameters (depending on the number of years of the simulation, T) are shown.

| Model | Seasonal | Interannual | Model equation | Parameters | Number of parameters |
|---|---|---|---|---|---|
| **A. Continuous and smooth interannual changes and periodic seasonality** | | | | | |
| A.0 | None | None | $\log X(t) = \log \bar{x}$ | $\bar{x}$ | 1 |
| A.1 | Periodic | None | $\log X(t) = \log \bar{x} + A \sin 2\pi d(t-a)$ | $\bar{x}, A, d, a$ | 1 + 3 |
| A.2 | None | Spline | $\log X(t) = \log \bar{x} + \text{spline}(t \mid x_1, \dots, x_{T+1})$ | $\bar{x}, x_1, \dots, x_{T+1}$ | 1 + (T+1) |
| A.3 | Periodic | Spline | $\log X(t) = \log \bar{x} + \text{spline}(t \mid x_1, \dots, x_{T+1}) + A \sin(2\pi d(t-a))$ | $\bar{x}, A, d, a, x_1, \dots, x_{T+1}$ | 1 + (T+1) + 3 |
| **B. Discontinuous interannual changes with aperiodic seasonality** | | | | | |
| B.0 | None | None | $\log X(t) = \log \bar{x}$ | $\bar{x}$ | 1 |
| B.1 | Pattern | None | $\log X(t) = \log \bar{x} + x_{m(t)}$ | $\bar{x}, x_{m_1}, \dots, x_{m_{12T}}$ | 1 + 12T |
| B.2 | None | Deviates | $\log X(t) = \log \bar{x} + x_{y(t)}$ | $\bar{x}, x_{y_1}, \dots, x_{y_T}$ | 1 + T |
| B.3 | Pattern | Deviates | $\log X(t) = \log \bar{x} + x_{y(t)} + x_{m(t)}$ | $\bar{x}, x_{y_1}, \dots, x_{y_T}, x_{m_1}, \dots, x_{m_{12T}}$ | 1 + T + 12T |
| **C. Interannual variability as a short pulse** | | | | | |
| C.0 | None | Gaussian pulse | $\log X(t) = \log \bar{x} - \frac{(x-t_0)^2}{2\sigma^2} - \log 2\sigma$ | $\bar{x}, \sigma, t_0$ | 3 |

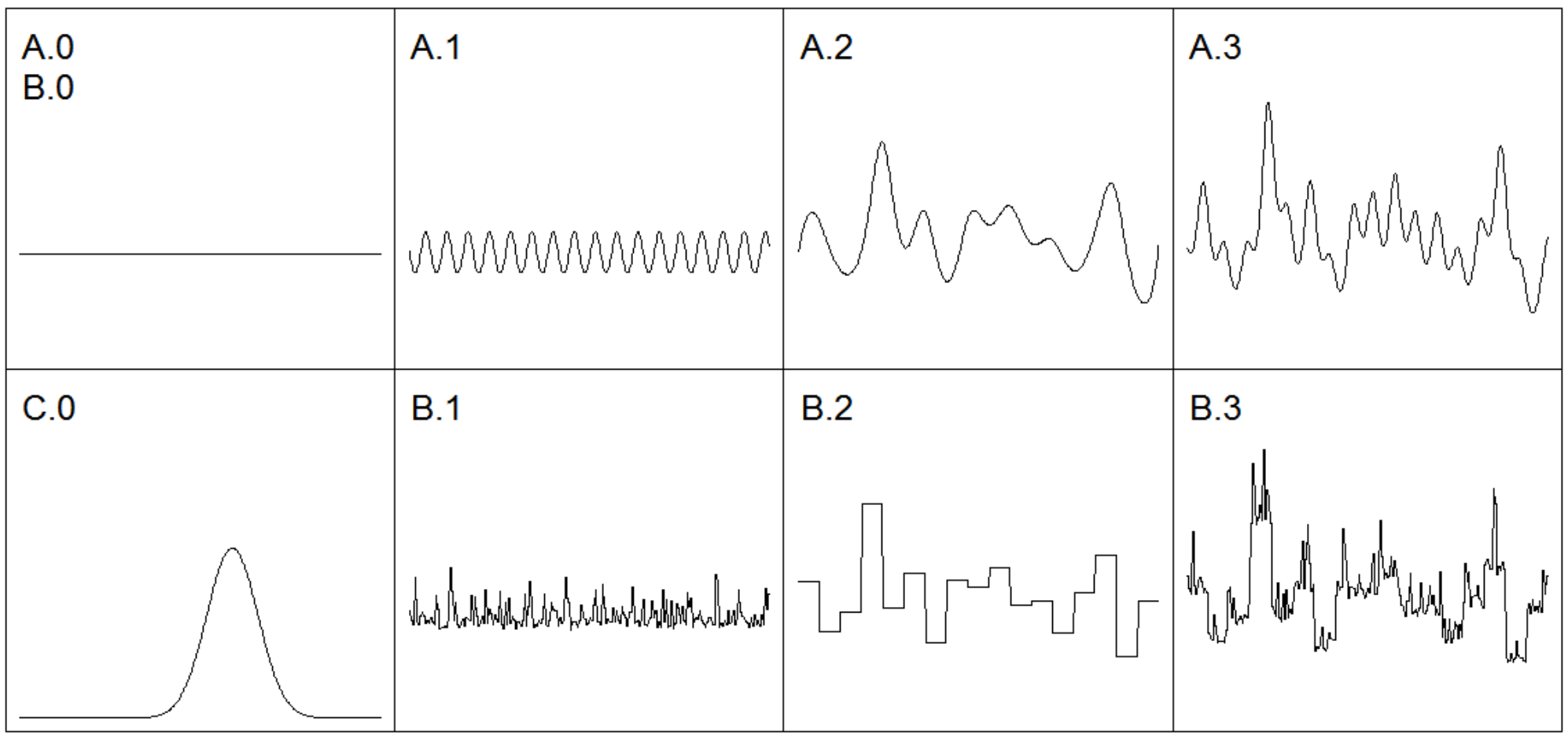

Figure B.1. Examples of the different models used for time-varying parameters.